\documentclass[12pt,preprint]{aastex}

\newcommand{\msun}{M$_{\odot}$}
\newcommand{\teff}{T$_{\mathrm{eff}}$}

\usepackage[T1]{fontenc}
\usepackage{ae,aecompl}

\usepackage{subfigure}
\usepackage{amsmath}
\usepackage{graphicx}
\usepackage{tablefootnote}
\usepackage{float}
\usepackage{url}
\usepackage{natbib}
\usepackage{amssymb}
\usepackage{color}
\usepackage{dblfloatfix}
\usepackage{fixltx2e}
\usepackage{morefloats}
\usepackage{appendix}

\usepackage[bookmarks=false]{hyperref}

\begin{document}

\title{DA White Dwarfs in the \emph{Kepler} Field}

\author{T.~F.~Doyle\altaffilmark{1,4}, S. B. Howell\altaffilmark{2,4,5}, V. Petit\altaffilmark{1}, and S. L\'epine\altaffilmark{3}}

\altaffiltext{1}{Florida Institute of Technology, 150 W. University Blvd., Melbourne, FL, 32907, USA}
\altaffiltext{2}{NASA Ames Research Center, Moffett Field, Mountain View, CA, 94035, USA}
\altaffiltext{3}{Georgia State University, 33 Gilmer Street SE, Atlanta, GA, 30302, USA}
\altaffiltext{4}{Visiting astronomer at Palomar Observatory}
\altaffiltext{5}{Visiting astronomer, Kitt Peak National Observatory, National Optical Astronomy Observatory, which is operated by the Association of Universities for Research in Astronomy (AURA) under a cooperative agreement with the National Science Foundation.}

\date{Accepted 2016 September 28. Received 2016 September 27; in original form 2016 June 21}

\begin{abstract}
We present 16 new, and confirm 7 previously identified, DA white dwarfs in the \emph{Kepler} field through ground-based spectroscopy with the Hale 200$''$, Kitt Peak 4-meter, and Bok 2.3-meter telescopes.  Using atmospheric models we determine their effective temperatures and surface gravities to constrain their position with respect to the ZZ Ceti (DA pulsator) instability strip, and look for the presence or absence of pulsation with \emph{Kepler's} unprecedented photometry.  Our results are as follows: \\
i) From our measurements of temperature and surface gravity, 12 of the 23 DA white dwarfs from this work fall well outside of the instability strip.  The \emph{Kepler} photometry available for 11 of these WDs allows us to confirm that none are pulsating.  One of these eleven happens to be a presumed binary, KIC 11604781, with a period of $\sim$5 days.  \\
ii) The remaining 11 DA white dwarfs are instability strip candidates, potentially falling within the current, empirical instability strip, after accounting for uncertainties.  These WDs will help constrain the strip's location further, as eight are near the blue edge and three are near the red edge of the instability strip.  Four of these WDs do not have \emph{Kepler} photometry, so ground-based photometry is needed to determine the pulsation nature of these white dwarfs.  The remaining seven have \emph{Kepler} photometry available, but do not show any periodicity on typical WD pulsation timescales.  

\end{abstract}


\section{Introduction}
White dwarfs (WDs) are extremely compact objects, and are the final evolutionary stage of $\sim$95\% of stars.  Even though WDs are well characterized and heavily studied, more can be learned about their internal structure through asteroseismology, especially since the advent of precise space-based photometry such as that provided by \emph{Kepler}.  We focus here specifically on characterizing DA white dwarfs located in the \emph{Kepler} field, and on constraining the onset of the instability strip.  

DA white dwarfs are the most common WD spectral type and have hydrogen-dominated envelopes (or photospheres), which produce strong hydrogen Balmer absorption lines.  Some DA white dwarfs are observed to pulsate.  For these WDs, asteroseismology can be used to determine their interior structure and composition (e.g., \citealt{gilliland10}).  Pulsating DA white dwarfs are found to exist only in a narrow region in the \teff-log $g$ plane, known as the instability strip.  The instability strip in DA white dwarfs is located in the temperature range 10 800 K $<$ \teff\ $<$ 12 300 K \citep{bergeron04, mukadam04}, where the $\kappa$- and $\gamma$-mechanisms in the hydrogen partial ionization zone drive the pulsations.  The instability strip was determined from observations to have a small dependence on mass as well (e.g., \citealt{giovannini98}).  

The purity of the instability strip for DA white dwarfs was first theorized by \citet{fontaine82}; if a DA white dwarf was found to lie within the instability strip, it would have to be pulsating.  This led to studies by \citet{fontaine82} and \citet{greenstein82} arguing that DA white dwarf pulsations are a phase through which all DA white dwarfs evolve as they cool.  If this is true, then studying the seismological properties of pulsating DA white dwarfs would provide constraints on the properties of all DA white dwarfs (e.g., \citealt{daou90}).  From theoretical modeling, the blue edge of the instability strip is predicted to be a sensitive function of WD mass and of the physical characteristics of the hydrogen envelope, especially the effectiveness of convective mixing \citep{wingfont82}.  Studying the blue edge of the instability strip by determining effective temperatures of pulsating, and non-pulsating, DA white dwarfs which fall near this boundary will inform us about the pulsation mechanism and its characteristics.  Additionally, studying the red edge of the instability strip allows for the determination of which mechanism is responsible for stopping pulsations at this point \citep{tassoul90,gianninas05}.  

In this paper, we set out to find pulsating DA white dwarfs in the \emph{Kepler} field.  \emph{Kepler} photometry is ideal for measuring photometric variability because of its precision and long timeline of nearly constant observations.  Modeling and analysis of spectroscopic observations of WDs allow for the determination of effective temperature and surface gravity.  These properties are used to determine whether a DA white dwarf falls within the instability strip, and whether it should thus be pulsating.  Photometric observations are then used to verify the hypothesis of pulsation, or stability, when \emph{Kepler} data is available.  

We present spectroscopic and photometric data for 23 DA white dwarfs in the \emph{Kepler} field.  We discuss the identification and selection of WD candidates, as well as details of the observations and data reduction, in \S\ref{sec:data}.  We then discuss the spectral modeling used to determine each white dwarf's effective temperature (\teff) and surface gravity (log $g$) and period analysis in \S\ref{sec:model}.  Results of the modeling and period analysis are presented in \S\ref{sec:dawds} for each of the 23 DA white dwarfs.  Finally, the conclusions are presented in \S\ref{sec:conclusion}.

\section{Observations and Data Reduction}\label{sec:data}
Target objects were chosen based on a program which surveyed the \emph{Kepler} field in $UBV$ in order to find new blue (hot) objects.  From this survey \citep{everett12}, photometric calibrators in the form of hot WDs were searched for, as non-pulsating WDs are known to be very stable photometric sources.  We searched the photometric data for objects with colors near $B-V=0.0$ and $U-B=-0.8$ or bluer, yielding several hundred candidates.  Candidates for the 2.3-meter Bok telescope follow-up program (see 2.1 below) were selected based on their blue color alone. Candidates for the Hale 200$''$ follow-up program  were selected based on a combination of blue color and high proper motion, performed as follows.  The list of blue candidates was cross-correlated against the SUPERBLINK all-sky proper motion catalog (\citealt{lepine11}; L\'epine, in prep.), which identifies stars with proper motions, $\mu>40$ mas yr$^{-1}$ and visual magnitudes $V$ brighter than 20, including in the \emph{Kepler} field. The identification of a faint blue object as a high proper motion star usually indicates that it is a hot subdwarf or white dwarf  \citep{lanning06}.  Our subset of $<$100 blue, high proper motion stars, along with some additional blue objects, were observed using ground-based optical spectroscopy (discussed in the following section) to determine their spectral types.  The ground-based spectroscopy revealed various spectral types of WDs, active galactic nuclei, and chromospherically active stars, among others.  Four sources (KIC 4829241, 6212123, 3354599, and 10149875) were selected as likely hot WDs in the \emph{Kepler} input catalog after being matched to sources on the POSS I (Palomar Observatory Sky Survey) blue plates.  

White dwarfs with counterparts in the SUPERBLINK proper motion catalog are denoted below with their SUPERBLINK catalog number (``PMI'' prefix) in addition to their KIC number. Other white dwarfs are identified by their catalog number from the $UBV$ survey (``Blue'' prefix) and from the Howell-Sing-Holberg Survey, conducted on the Bok 2.3-meter telescope (``HSH'' prefix).  As the HSH spectra turned out to be of low quality, the subsequent observations were obtained with the 4-meter telescope.

\subsection{Spectroscopic Observations}\label{sec:palomar}
Optical, ground-based spectroscopic observations were obtained for 10 of the WDs from the Palomar 200$''$ (5.1-meter) Hale Telescope on Mount Palomar in California.  We used the Double Beam Spectrograph (DBSP), which is composed of two channels, blue and red, with wavelength ranges of 3500\AA\ - 5000\AA\ and 5500\AA\ - 7500\AA, respectively.   The observations utilized blazed gratings with low to medium resolution, R $\approx$ 4500 (red channel) and R $\approx$ 3000 (blue channel), and a slit width of $\sim$1$''$.  Observations were taken over several observing runs throughout August 2013.  For 9 of the WDs, optical, ground-based observations were obtained from the Kitt Peak Mayall 4-meter telescope using the KPC-22b grating in the second order on the Ritchey Chr\'etien (RC) spectrograph.  The blue spectral resolution for the setup used is $\sim$5000, providing a wavelength coverage of 3700-5100\AA\ with a dispersion on the CCD detector of 0.72\AA\ pixel$^{-1}$.  The slit was set to 1$''$ and used in an east-west (90$^{\circ}$) orientation for all observations. For the 4 remaining WDs, optical, ground-based observations were obtained with the Steward Observatory 2.3-meter Bok telescope at Kitt Peak, using the Boller and Chivens spectrograph.  The 2005 September observations were taken with the 832 lines/mm grating and used in the 2nd order at two different grating tilts to cover the blue region ($\sim$3400-4850\AA) and 1st order in conjunction with a UV filter to cover the red region (5000-7000\AA) with two separate grating tilts. A 1.5$''$ slit was used producing a resolving power of $\sim$2700.  An observing log for all of our spectroscopy is presented in Table~\ref{table:wdobs}.  

All spectra were reduced with \textsc{iraf} (Image Reduction and Analysis Facility) software, using well-known routines (\textsc{noao/imred} package) to extract one-dimensional optical spectra.  A single one-dimensional spectrum is extracted from the raw spectrum by determining the extraction region, subtracting out the background, and fitting a function along the extraction axis.  Once the spectrum is extracted, it is then flux calibrated by applying a standard star to the spectrum.  It is then wavelength calibrated using arc or comparison lamp spectra, which have known rest wavelengths.  Blue and red spectra from the Hale telescope were reduced separately due to the gap in data and the difference in resolution between the channels; the channels were combined for analysis.  The blue channel spectra were analyzed first in order to identify the DA white dwarfs because all of the Balmer lines except H$\alpha$ are located within the blue wavelength range.

\begin{table*}[h]
\tiny
\centering
\caption{Spectroscopic observations of  the DA white dwarfs.}
\begin{tabular}{l c c c c c c c c c}
\hline
\hline
\textbf{Object Name} & \textbf{KIC} & \textbf{Obs Date} & \textbf{$V$} & \textbf{$B-V$} & \textbf{$U-B$} & \textbf{\# Exposures} & \textbf{Exp Time} & \textbf{Channel} & \textbf{Telescope} \\ 
 &  & DD-MM-YYYY & mag & mag & mag & per channel & s &  \\ 
\hline
PMI18450+4800 & 10709534 & 10-08-2013 & 18.7 & 0.2 & -0.4 & 2 & 900 & red/blue & Hale \\
PMI18486+4811 & 10777440 & 10-08-2013 & 18.5 & 0.9 & $\cdots$ & 2 & 900 & red/blue & Hale \\
PMI18553+4755 & 10649118 & 09-08-2013 & 18.8 & 0.3 & -0.6 & 2 & 900 & red/blue & Hale \\
PMI19002+3922 & 4242459 & 09-08-2013 & 15.7 & 0.2 & -0.5 & 1 & 900 & red/blue & Hale \\
PMI19085+4338 & 7879431 & 11-08-2013 & 18.1 & 0.1 & -0.7 & 2 & 600 & red/blue & Hale \\
PMI19141+4936 & 11604781 & 11-08-2013 & 16.7 & 0.4 & -0.7 & 2 & 600 & red/blue & Hale \\
PMI19173+4452 & 8682822 & 11-08-2013 & 15.7 & 0.0 & -0.9 & 2 & 600 & red/blue & Hale \\
PMI19179+4524 & 9082980 & 11-08-2013 & 18.3 & 0.1 & -0.6 & 2 & 750 & red/blue & Hale \\
PMI19245+3734 & 2158770 & 11-08-2013 & 18.2 & 0.6 & 0.4 & 2 & 900 & red/blue & Hale \\
PMI19409+4240 & 7129927 & 09-08-2013 & 16.5 & 0.3 & -0.6 & 1 & 600 & red/blue & Hale \\
Blue18 & 6042560 & 01-06-2012 & 16.7 & $\cdots$ & $\cdots$ & 1 & 1200 & blue & Mayall \\
Blue1903 & 7346018 & 01-08-2011 & 16.7 & 0.0 & -1.0 & 1 & 1200 & blue & Mayall \\
PMI19010+4208 & 6672883 & 01-06-2012 & 17.4 & 0.1 & -0.5 & 1 & 1200 & blue & Mayall \\
PMI19060+4446 & 8612751 & 01-06-2012 & 17.3 & 0.3 & -0.5 & 1 & 1200 & blue & Mayall \\
PMI19099+4717 & 10198116 & 01-06-2012 & 16.3 & 0.1 & -0.5 & 1 & 1200 & blue & Mayall \\
PMI19320+4925 & 11509531 & 01-06-2012 & 17.7 & 0.2 & -0.6 & 1 & 1200 & blue & Mayall \\
PMI19362+4714 & 10213347 & 01-06-2012 & 18.0 & 0.2 & -0.6 & 1 & 1200 & blue & Mayall \\
PMI19423+4407 & 8244398 & 01-06-2012 & 17.6 & 0.2 & -0.5 & 1 & 1200 & blue & Mayall \\
PMI19430+4538 & 9228724 & 01-06-2012 & 17.3 & 0.1 & -0.5 & 1 & 1200 & blue & Mayall \\
HSH08 & 4829241 & 05-09-2005 & 15.7 & 0.0 & -0.8 & 1 & 600 & red/blue & Bok \\
HSH24 & 6212123 & 05-09-2005 & 16.6 & -0.2 & -1.1 & 1 & 600 & red/blue & Bok \\
HSH32 & 3354599 & 05-09-2005 & 16.7 & 0.1 & -0.9 & 1 & 500 & red/blue & Bok \\
HSH36 & 10149875 & 05-09-2005 & 17.7 & -0.2 & -1.2 & 1 &  1200 & red/blue & Bok \\
\hline
\end{tabular}
\label{table:wdobs}
\end{table*}

\subsection{Photometric Observations}\label{sec:kepler}
Photometric observations were obtained by the \emph{Kepler} 0.95m space telescope, which has a bandpass of 4300-8900\AA\footnote{\url{http://kepler.nasa.gov/}}.  \emph{Kepler} observed a 115 square-degree field of view in the Cygnus constellation almost non-stop, obtaining unprecedented photometric coverage of this field from 2009-2013.  Discovering planets via transits was the main objective of \emph{Kepler}, but its photometric precision and depth, down to 21 \emph{Kepler} magnitudes, is also extremely beneficial for astronomical studies of variable systems.  \emph{Kepler} has two modes of observation, long cadence (LC; 30 minute exposures) and short cadence (SC; 1 minute exposures), both of which are used in this work.    All \emph{Kepler} data are archived and publicly available from the Mikulski Archive for Space Telescopes\footnotemark[1] (MAST), where the light curves for our DA white dwarfs were obtained.  Light curves were available for eighteen of the twenty three DA white dwarfs.  

There are two pipeline-reduced data sets available for download and analysis, PDCSAP\footnote{\url{archive.stsci.edu/kepler/manuals/archive_manual.pdf}} (presearch data conditioning simple aperture photometry) and SAP (simple aperture photometry).  Both of these pipeline-reduced light curves are adequate for photometric studies, although the user can obtain the raw data and re-reduce it using \textsc{pyraf} routines, if necessary.  PDCSAP data was utilized for the DA white dwarfs in this paper.  Table~\ref{table:dawdlc} lists the photometric observation information for all of our white dwarfs in columns 8 and 9.  LC and SC refer to long and short cadence data and ``Qtrs'' indicates the appropriate 90 day blocks of \emph{Kepler} data.

\begin{table*}[h]
\tiny
\centering
\caption{Modeled parameters and \emph{Kepler} photometric data of the 23 DA white dwarfs.}
\begin{tabular}{l c c c c c c c c c}
\hline
\hline
\textbf{KIC} & \textbf{RA} & \textbf{DEC} & \textbf{T$_{\textbf{eff}}$} & \textbf{$\sigma_{T_{\mathrm{eff}}}$} & \textbf{log $g$} & \textbf{$\sigma_{\mathrm{log\ g}}$} & \textbf{Qtrs} & \textbf{Qtrs} & \textbf{Instability Strip} \\
 & HH:MM:SS.ss & DD:MM:SS.ss & kK & kK & dex &  & LC & SC & \textbf{Candidate?}  \\ \hline
10709534 & 18:45:05.63 & +48:00:41.20 & 10.5 & $^{+3.4}_{-1.0}$ & 8.50 & $^{+0.43}_{-0.60}$ & -- & -- & yes \\
10777440 & 18:48:37.94 & +48:11:24.93 & 14.8 & $^{+1.9}_{-2.4}$ & 8.25 & $^{+0.33}_{-0.35}$ & -- & -- & yes \\
10649118 & 18:55:21.39 & +47:55:44.94 & 8.50 & $^{+0.82}_{-0.69}$ & 8.00 & $^{+0.64}_{-0.77}$ & -- & -- & no \\
4242459 & 19:00:17.34 & +39:22:32.90 & 9.50 & $^{+0.22}_{-0.13}$ & 8.25 & $^{+0.09}_{-0.05}$ & 1-17 & 6-13 & no \\
7879431 & 19:08:34.01 & +43:38:30.16 & 14.5 & $^{+2.2}_{-2.3}$ & 8.25 & $^{+0.32}_{-0.42}$ & -- & -- & yes \\
11604781 & 19:14:08.99 & +49:36:41.04 & 9.50 & $^{+0.55}_{-0.26}$ & 8.50 & $^{+0.32}_{-0.10}$ & 3,6,7,10,11 & 3 & no \\ 
8682822 & 19:17:20.59 & +44:52:39.78 & 19.5 & $^{+1.3}_{-1.1}$ & 8.75 & $^{+0.14}_{-0.10}$ & 1,5-9 & 1,5 & no \\
9082980 & 19:17:58.06 & +45:24:27.40 & 14.0 & $^{+1.9}_{-1.9}$ & 8.00 & $^{+0.42}_{-0.27}$ & -- & -- & yes \\
2158770 & 19:24:33.02 & +37:34:16.57 & 12.5 & $^{+3.3}_{-1.9}$ & 8.00 & $^{+0.46}_{-0.50}$ & 6-9 & -- & yes \\
7129927 & 19:40:59.38 & +42:40:31.48 & 9.50 & $^{+0.45}_{-0.31}$ & 8.25 & $^{+0.30}_{-0.15}$ & 3,5,6 & 3 & no \\ 
6042560 & 19:27:28.52 & +41:20:08.83 & 6.75 & $^{+0.76}_{\ge-0.75}$ & 7.75 & $^{+1.11}_{\ge-0.75}$ & 16,17 & -- & no \\
7346018 & 19:03:17.27 & +42:56:49.86 & 30.0\footnotemark[4]\footnotetext[4]{upper limit} & $^{\ \ \cdots}_{-3.5}$ & 8.50 & $^{+0.50}_{-0.82}$ & 16,17 & -- & no \\
6672883 & 19:01:02.29 & +42:08:39.03 & 16.5 & $^{+6.6}_{-6.8}$ & 7.75 & $^{+1.14}_{\ge-0.75}$ & 16 & -- & yes \\
8612751 & 19:06:05.04 & +44:46:59.98 & 7.50 & $^{+1.35}_{-1.08}$ & 8.00 & $^{\ge+1.00}_{\ge-1.00}$ & 16,17 & -- & no \\
10198116 & 19:09:59.36 & +47:17:10.41 & 13.5 & $^{+3.4}_{-2.1}$ & 8.00 & $^{+0.35}_{-0.31}$ & 4-6 & 4 & yes \\
11509531 & 19:32:01.31 & +49:25:32.99 & 9.25 & $^{+5.36}_{-1.52}$ & 8.25 & $^{\ge+0.75}_{\ge-1.25}$ & 16,17 & -- & yes \\
10213347 & 19:36:12.14 & +47:14:22.37 & 10.3 & $^{+17.6}_{-2.0}$ & 8.00 & $^{\ge+1.00}_{\ge-1.00}$ & 16,17 & -- & yes \\
8244398 & 19:42:18.45 & +44:07:45.66 & 13.5 & $^{+7.3}_{-3.5}$ & 8.00 & $^{+0.76}_{-0.90}$ & 16,17 & -- & yes \\
9228724 & 19:43:02.56 & +45:38:42.57 & 12.8 & $^{+6.8}_{-2.3}$ & 8.25 & $^{+0.49}_{-0.88}$ & 16,17 & -- & yes \\
4829241 & 19:19:27.68 & +39:58:39.25 & 19.5 & $^{+1.5}_{-1.6}$ & 8.00 & $^{+0.29}_{-0.24}$ & 1-13 & 1,5 & no \\
6212123 & 19:36:04.94 & +41:33:06.07 & 6.25 & $^{+0.59}_{\ge-0.25}$ & 8.50 & $^{\ge+0.50}_{-0.82}$ & 2-13 & -- & no \\
3354599 & 19:38:05.78 & +38:25:30.04 & 6.00\footnotemark[5]\footnotetext[5]{lower limit} & $^{+0.92}_{\ \ \cdots}$ & 8.50 & $^{+0.50}_{-1.09}$ & 2-9, 11-13 & -- & no \\
10149875 & 19:40:13.36 & +47:09:48.78 & 30.0\footnotemark[4] & $^{\ \ \cdots}_{-0.8}$ & 9.00\footnotemark[4] & $^{\ \ \cdots}_{-0.29}$ & 16,17 & -- & no \\
\hline
\end{tabular}
\label{table:dawdlc}
\end{table*}

\section{Modeling and Analysis}\label{sec:model}
Each DA white dwarf was initially identified visually in the spectroscopic data from its easily recognizable spectral signature: DA white dwarf spectra exhibit characteristic, gravity-broadened absorption lines of the hydrogen Balmer series.  The parameters of the DA white dwarfs were then determined by fitting atmospheric models to the observed spectra.  Based on the derived effective temperature and surface gravity from the best fit model, we assessed whether the WD would be expected to be in the instability strip, and thus a pulsator.  We then used the \emph{Kepler} photometric data, when available, to determine if there is any evidence of variability and to confirm or deny our pulsation hypothesis.   

\subsection{Spectral Models}\label{sec:specmodels}
DA white dwarf models were provided by Detlev Koester (priv. comm.); a discussion of the models and their usage can be found in \citet{koester10}.  The models were produced using four basic assumptions of modeling stellar atmospheres: 1) homogeneous, plane-parallel geometry, 2) hydrostatic equilibrium, 3) radiative and convective equilibrium, and 4) local thermodynamic equilibrium.  We used a group of 594 models with a temperature grid spacing of 250 K from 6000 K to 20 000 K and 1000 K from 20 000 K to 30 000 K, with log $g$ spacing of 0.25 dex from 7.00 to 9.00 dex.  Errors for measured effective temperature and gravity at the edges of the model range are either marked as zero (in the text), as ``$\cdots$'' (in Table~\ref{table:dawdlc}), or left blank (in Figs.~\ref{fig:blue1903plots}, \ref{fig:hsh32plots}, and \ref{fig:hsh36plots}).  

In order to fit the models to the observed DA white dwarf spectra, both the models and the spectra were normalized to the continuum.  For all lines (H$\alpha$ to H$\xi$), except H$\beta$ (for the Hale telescope), a continuum section on either side of the spectral line was selected and a linear regression was fit to these sections, excluding the absorption line itself.  For H$\beta$ (Hale only), only the continuum on the blue side of the line was used, as the red wing of H$\beta$ is diminished in intensity due to the grating efficiency falling off on the red side of H$\beta$.  There is no systematic difference between the fits of H$\alpha$ alone versus the inclusive fits of H$\beta$-H$\xi$ and H$\delta$-H$\xi$ in the Hale spectra.  This process was used for the spectral models as well, to normalize the modeled and observed spectral continuum to unity.  The models were folded with a Gaussian profile matching the instrumental width of the spectrograph before being fitted to the observed spectra.  

Each Balmer line was then fitted individually for each WD, and the lowest $\chi^2$ value was calculated.  The best global fit was determined by fitting all the lines simultaneously and minimizing the total $\chi^2$ for all five (or six) lines (H$\beta$ to H$\xi$ and H$\alpha$ where available).  Stacked plots of all the Balmer lines and their best global model fits are discussed and displayed in \S\ref{sec:dawds}.  Also displayed in \S\ref{sec:dawds} are $\chi^2$ contour maps in \teff-log $g$ space, which show the best fit model, the lowest $\chi^2$, and the empirical instability strip for comparison.  A discussion of the errors and how they were determined can also be found in \S\ref{sec:dawds}.  

\subsection{Period Analysis}\label{sec:period}
\emph{Kepler} photometric data of each WD was run through an \textsc{idl} routine in order to compute the Lomb-Scargle periodogram.  Lomb-Scargle is a mathematical computation, similar to a discrete Fourier transform (DFT), which calculates the most likely period of time series data \citep{scargle82}.  These calculations allowed us to determine if there were any likely periods of the WDs, which would imply pulsations.  Pulsating DA white dwarfs typically have periods of up to $\sim$1 day, but usually much less than 1 day, $\sim$100 to 1000 s (e.g., \citealt{bergeron95}, \citealt{fontbrass08}).  
The range searched when looking for periodic trends in the \emph{Kepler} data was 0.001 to 20.0 days.  Pulsation periods were searched for in the 0.001 to $\sim$1.0 day range, whereas above $\sim$1.0 day was the region where we searched for any other type of variability (i.e., companions).  The low end of the pulsation search range goes much lower than the normal range of pulsations, but it was extended in order to be as inclusive as possible.  

Calculations of likely periods were also conducted using the Exoplanet Archive\footnote{\url{http://exoplanetarchive.ipac.caltech.edu/}} periodogram tool, which resulted in the same results as the above analysis.  

Eighteen of the twenty three DA white dwarfs have \emph{Kepler} light curves in the archive and the data are discussed in \S\ref{sec:dawds}.  Two of the DA white dwarfs with \emph{Kepler} data show variability, but both periods are not within the range of pulsations (see discussions in \S\ref{sec:pmi19141} and \ref{sec:hsh08}).  The remaining sixteen WDs with \emph{Kepler} light curves did not reveal any statistically significant ($>$3$\sigma$) periods within the normal range of WD pulsations.   

\section[]{Results}\label{sec:dawds}

\begin{figure*}
\centering
	\includegraphics[width=0.7\textwidth]{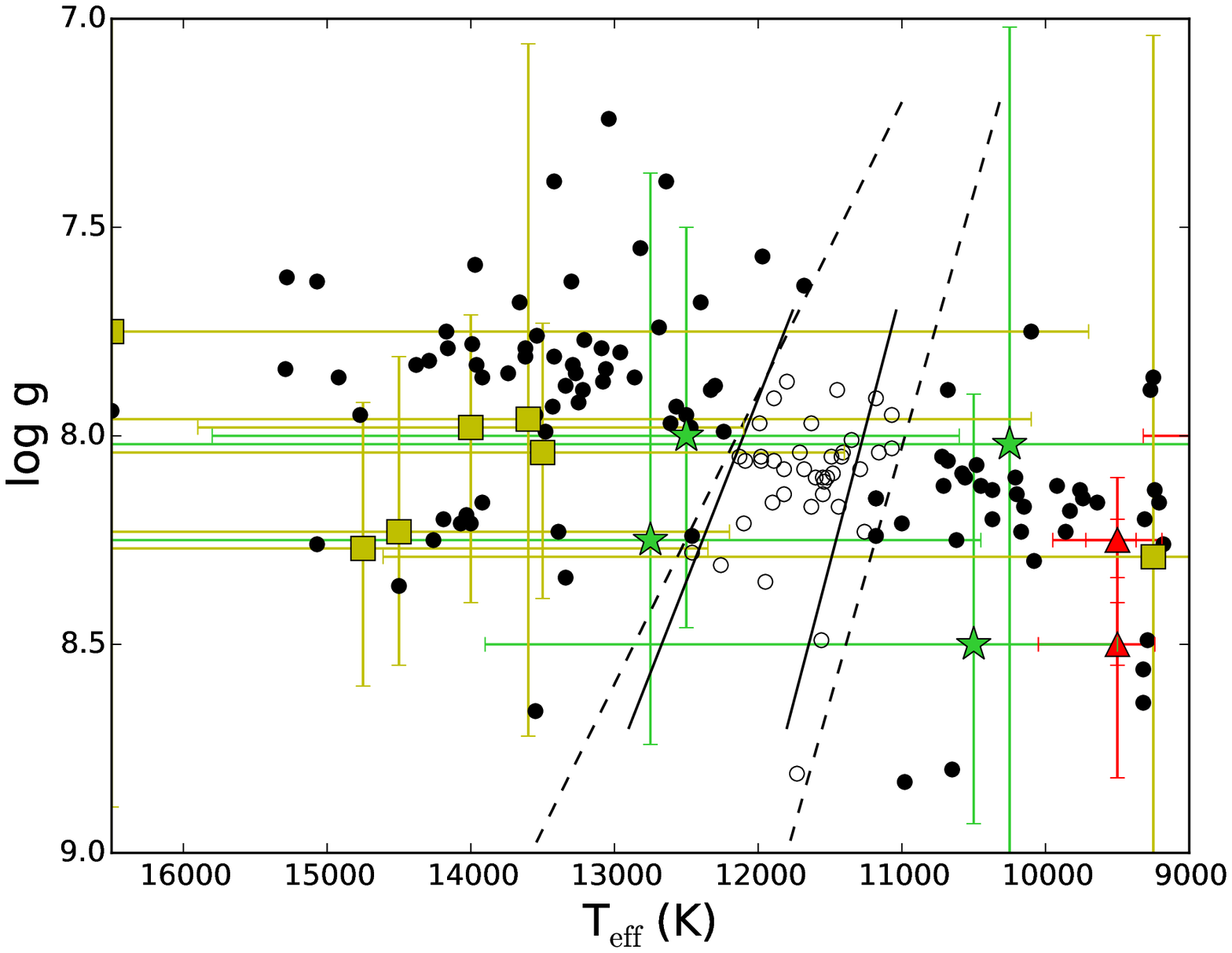}
    \caption{Reproduced figure from \citet{gianninas05} with the empirical (dashed line) and the theoretical (solid line) instability strip calculated from \citet{fontaine03}, known photometrically stable WDs (filled circles) from \citet{gianninas05} and known ZZ Cetis (open circles) from \citet{bergeron04}.  DA white dwarfs from this paper are displayed as colored shapes.  Only 13 of the 23 DA white dwarfs are shown in this figure; the remaining 10 are well outside the bounds of the instability strip.  Yellow squares and green stars indicate the eleven instability strip candidates.  The green stars are the four WDs that lie close enough to the instability strip that their status as instability strip candidates is more likely than the remaining seven WDs (yellow squares). Some of the instability strip candidates have been shifted in log $g$ by 0.02 or 0.04 dex and one shifted in \teff\ by 100 K to disentangle the error bars.  The red triangles are two of the twelve non-instability strip WDs which lie within this range of effective temperature and log $g$.}  
    \label{fig:instastrip}
\end{figure*}

Figure~\ref{fig:instastrip} shows the empirical instability strip in \teff-log $g$ space, plotted with known ZZ Cetis (open circles) and known photometrically stable DA white dwarfs (filled circles).  This figure is a reproduction of Fig. 6 from \citet{gianninas05}, with the addition of the DA white dwarfs from this paper (colored shapes).

In the following subsections, each DA white dwarf and its modeled parameters are discussed individually (a summary is provided in Table~\ref{table:dawdlc}).  Eleven of the twenty three DA white dwarfs are determined to be instability strip candidates based on their modeled spectroscopic parameters; seven of these WDs have \emph{Kepler} photometry.  Of the remaining twelve WDs, eleven fall well outside of the instability strip and can be added to the list of known photometrically stable WDs.  Ten of these WDs are confirmed as non-pulsators by \emph{Kepler} photometry and one does not have \emph{Kepler} photometry, but is not predicted to be an instability strip candidate.  The remaining WD, KIC 11604781, shows a clear periodicity near 5 days, attributed to an orbital variation.  Previous studies of this WD also conclude that it is probably a binary (see \S\ref{sec:pmi19141}).  We discuss the photometric data in this section and the confirmation, where possible, of our hypotheses of the pulsation nature of each WD from the spectral modeling described in \S\ref{sec:specmodels}.  

$\chi^2_{\nu}$ contour plots were created in order to determine the parameters' confidence intervals and statistical errors.  One, two and three $\sigma$ contours, which correspond to $\Delta\chi^2_{\mathrm{min}}+2.30$, $\Delta\chi^2_{\mathrm{min}}+6.18$, and  $\Delta\chi^2_{\mathrm{min}}+11.8$ \citep{numerical} respectively, are shown on each contour plot for each DA white dwarf, indicating the confidence intervals of the model fits.  The statistical errors of each best fit correspond to the 1$\sigma$ contour, where the upper and lower bounds of the contour at this level for both the gravity and effective temperature were determined.  H$\beta$ has a large effect on the errors as well because it is an upper Balmer series line, where the log $g$ effect is stronger than in the later lines.  The 4-meter data appear to have better S/N, even though the resolution is less than the Hale telescope; this is most likely due to observing conditions or greater integration time.  The ZZ Ceti empirical instability strip is shown on each contour map to illustrate its position with respect to the \teff-log $g$ confidence intervals.  If the 1$\sigma$ contour overlaps with the instability strip at any point, the WD is classified as an instability strip candidate.  

Seven of the twenty three DA white dwarfs in this sample have been previously modeled in the literature.   Previous parameter estimates are compared with parameters obtained from the present work in Table~\ref{table:comparison}.  Five of these seven WDs have previous spectral observations and analysis by \citet{ostensen10,ostensen11}.  These previous observations were conducted with multiple ground-based telescopes using low resolution spectra, with the resolving power, $R$, ranging from 550 to 1600 (see specifics in \citet{ostensen10,ostensen11}).  This resolving power is lower than that which we achieved with all three of the spectrographs used in this work.  Discrepancies arise in the determinations of effective temperature and gravity for these five WDs and are discussed in the appropriate sections below.  We note that the resolution for the Bok 2.3-m observations is lower than the Hale observations, therefore the model fits are not as good.  

\begin{table}[h]
\scriptsize
\centering
\caption{Comparison of This Work and the Literature.}
\begin{tabular}{l c c c c c}
\hline
\hline
\textbf{KIC} & \multicolumn{2}{c}{\textbf{This Work}} & \multicolumn{2}{c}{\textbf{Literature}} & \textbf{Reference} \\
 &  \teff & log $g$ &  \teff & log $g$ &  \\
 & kK & dex & kK & dex & \\
\hline
2158770 & 12.5 & 8.00 & 9.964 & 8.01 & \citealt{kleinman13} \\
4242459 & 9.50 & 8.25 & 9.470 & 8.00 & \citealt{zuckerman03} \\
11604781 & 9.50 & 8.50 & 9.1(5) & 8.3(3) & \citealt{ostensen11} \\
8682822 & 19.5 & 8.75 & 23.1 & 8.5 & \citealt{ostensen10} \\
7129927 & 9.50 & 8.25 & 23.314 & 7.280 & \citealt{ostensen11} \\
 &  &  & 24.191 & 7.120 & \citealt{ostensen11} \\
10198116 & 13.5 & 8.00 & 14.2(5) & 7.9(3) & \citealt{ostensen11} \\ 
4829241 & 19.5 & 8.00 & 19.4(5) & 7.8(3) & \citealt{ostensen11} \\ 
 &  &  & 20.376 & 7.93 & \citealt{zhao13} \\ 
\hline
\end{tabular}
\label{table:comparison}
\end{table}

\subsection{Instability Strip Candidates} \label{sec:puls}
\subsubsection{KIC 10709534 (PMI18450+4800)} \label{sec:pmi18450}
\begin{figure}[h]
    \begin{subfigure}[]{}
        \includegraphics[width=0.44\textwidth]{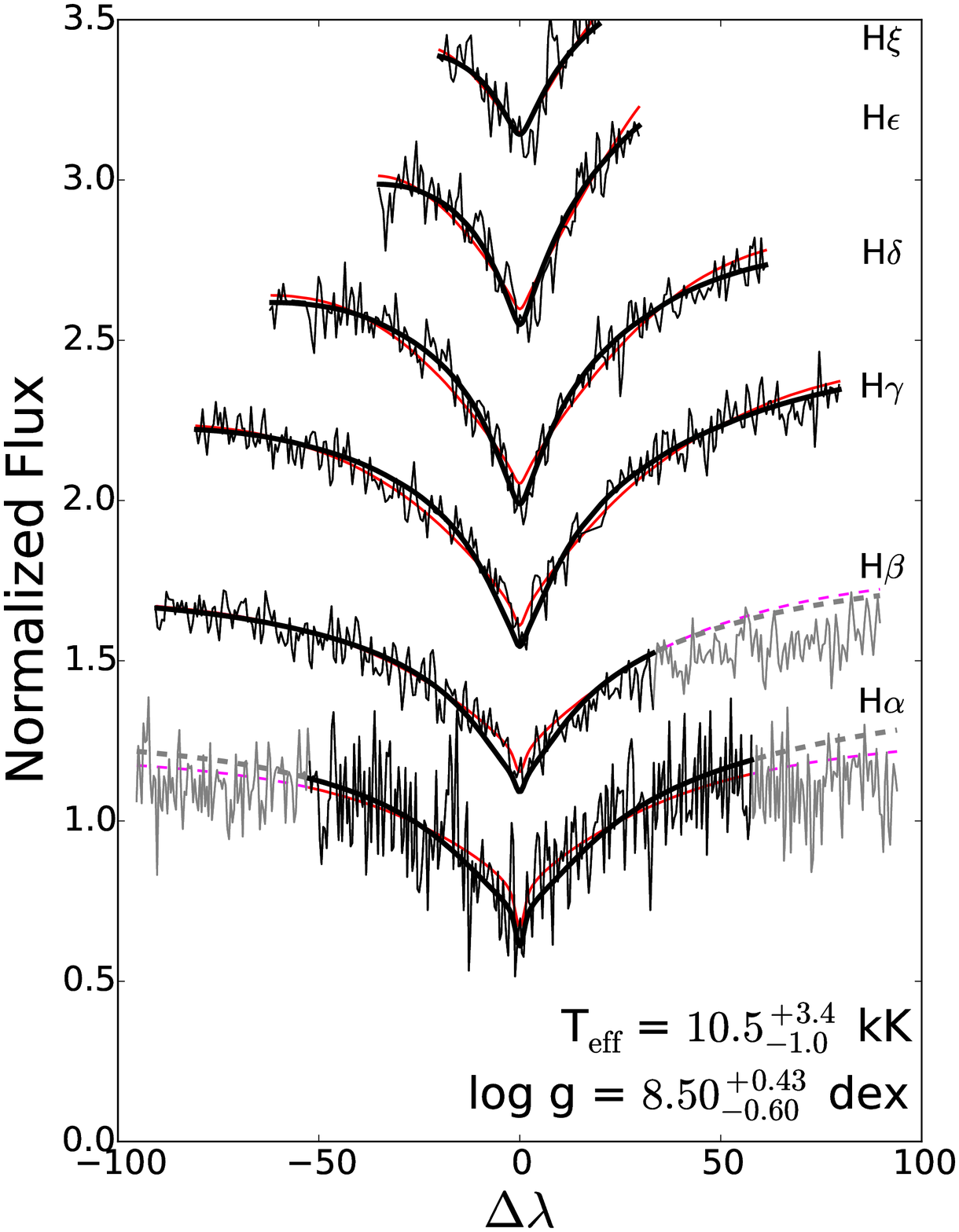}
    \end{subfigure}
    ~  
    \begin{subfigure}[]{}
        \includegraphics[width=0.47\textwidth]{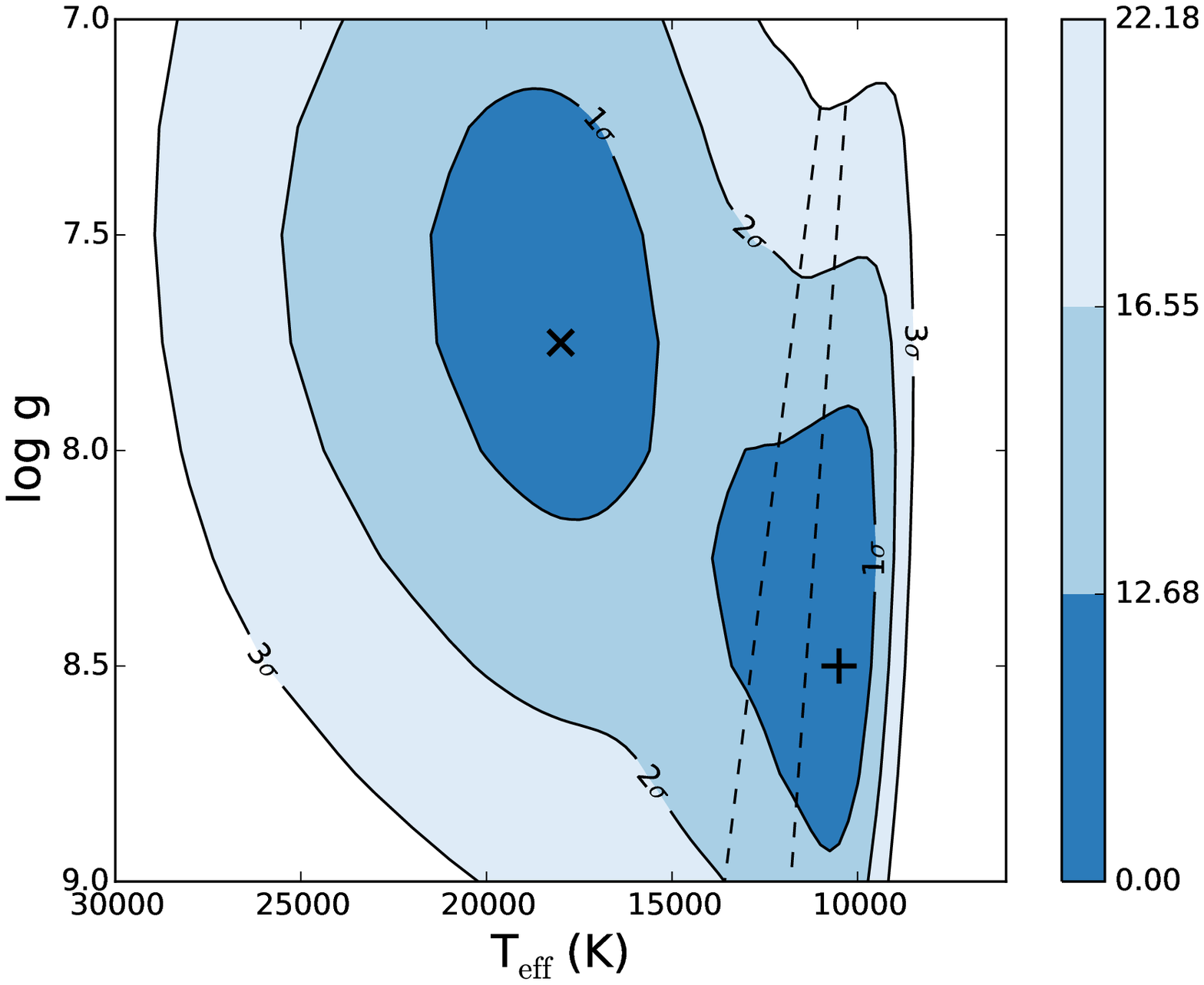}
    \end{subfigure}
    \caption{(a) Stacked Balmer line diagram of the best global model fit (thick black line) for KIC 10709534.  The red line fit corresponds to the secondary $\chi^2$ minimum at \teff\ = 18.0 kK and log $g$ = 7.75 dex.  From top to bottom, the corresponding spectral lines are H$\xi$, H$\epsilon$, H$\gamma$, H$\delta$, H$\beta$, and H$\alpha$.  (b) $\chi^2$ contour map for KIC 10709534 of all lines combined in \teff-log $g$ space.  The $+$ indicates the best fit in effective temperature and gravity.  The $\times$ corresponds to the secondary $\chi^2$ minimum fit.  The empirical instability strip is shown by the dashed line.}\label{fig:18450plots}
\end{figure}

KIC 10709534 is a new DA white dwarf in the \emph{Kepler} field with \teff\ = 10.5 $^{+3.4}_{-1.0}$ kK and log $g$ = 8.50 $^{+0.43}_{-0.60}$ dex.  Fig.~\ref{fig:18450plots}a shows the best fit model to each Balmer line (thick black line).  Fig.~\ref{fig:18450plots}b shows the location of the best fit (cross) in \teff-log $g$ space with contours of 1, 2 and 3$\sigma$.  The 1$\sigma$ error contour is shown in the \teff-log $g$ space to demonstrate that, taking into account the errors on both the effective temperature and surface gravity, KIC 10709534 has a high probability of being a pulsator.  Unfortunately, \emph{Kepler} photometric data for KIC 10709534 was unavailable, so the pulsation nature could not be confirmed.  

KIC 10709534's contour plot shows a double-minimum, allowing the effective temperature and surface gravity of this WD to be either the values stated above, or \teff\ = 18.0 $^{+3.5}_{-2.6}$ kK and log $g$ = 7.75 $^{+0.41}_{-0.59}$ dex.  If the modeled parameters were in fact the latter, KIC 10709534 would not be an instability strip candidate.  This double-minimum degeneracy shows that there are mathematically two distinct parameter subsets that provide statistically good fits to the spectrum.  The minimum $\chi^2$ of each 1$\sigma$ contour space in \teff-log $g$ space was however examined by eye, and the primary model value with \teff\ = 10.5 kK was determined to be a better fit of the Balmer line cores (see comparison of fits in Fig.~\ref{fig:18450plots}a). This is consistent with the marginally lower $\chi^2$ value found for this minimum, and suggests this value is the most likely estimate of the two.  In addition, $B-V$ and $U-B$ colors were used to determine an approximate effective temperature, which is most consistent with \teff\ = 10.5 kK.  \\

\clearpage 
\subsubsection{KIC 10777440 (PMI18486+4811)}
\begin{figure}[h]
    \begin{subfigure}[]{}
        \includegraphics[width=0.46\textwidth]{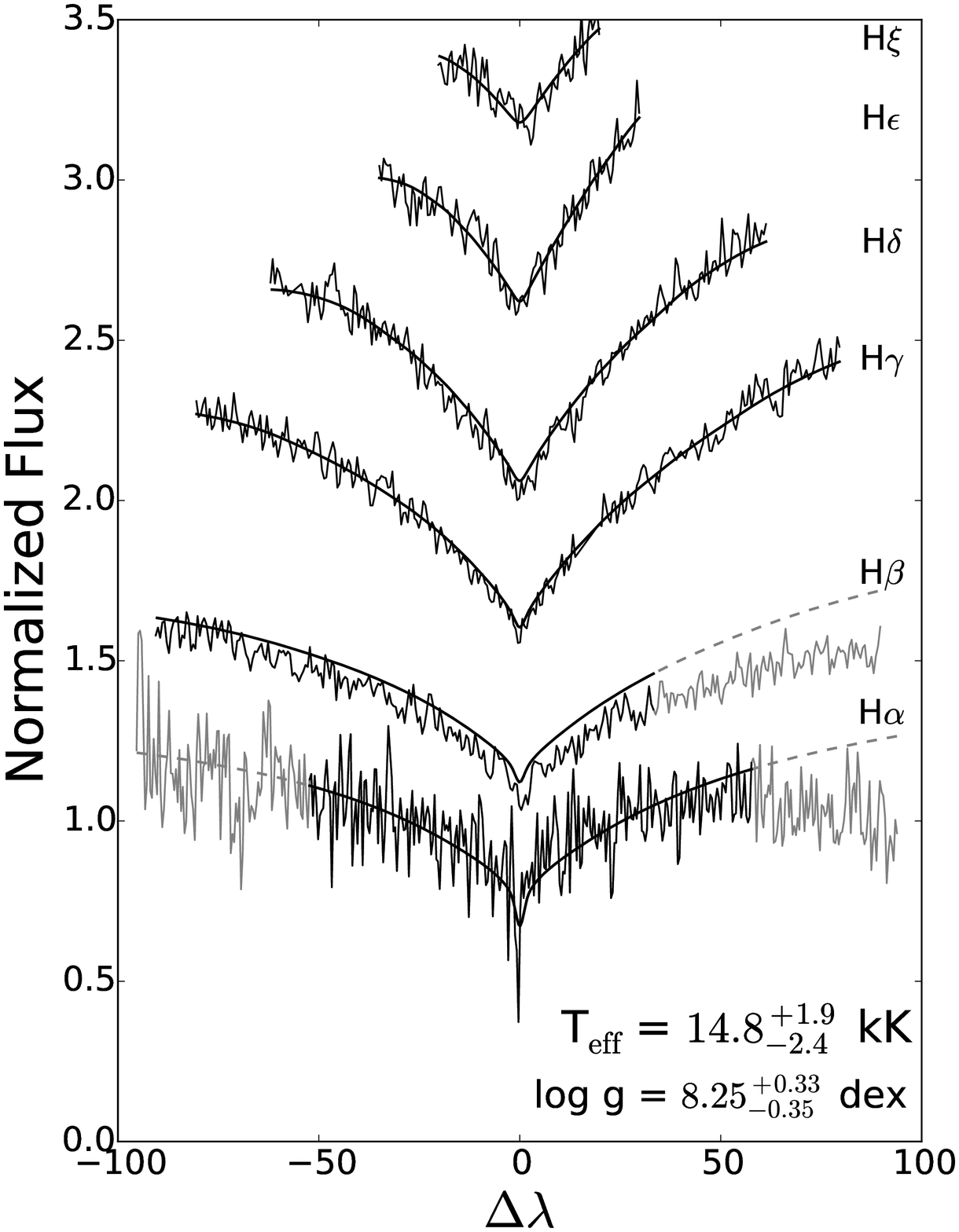}
    \end{subfigure}
    ~  
    \begin{subfigure}[]{}
        \includegraphics[width=0.47\textwidth]{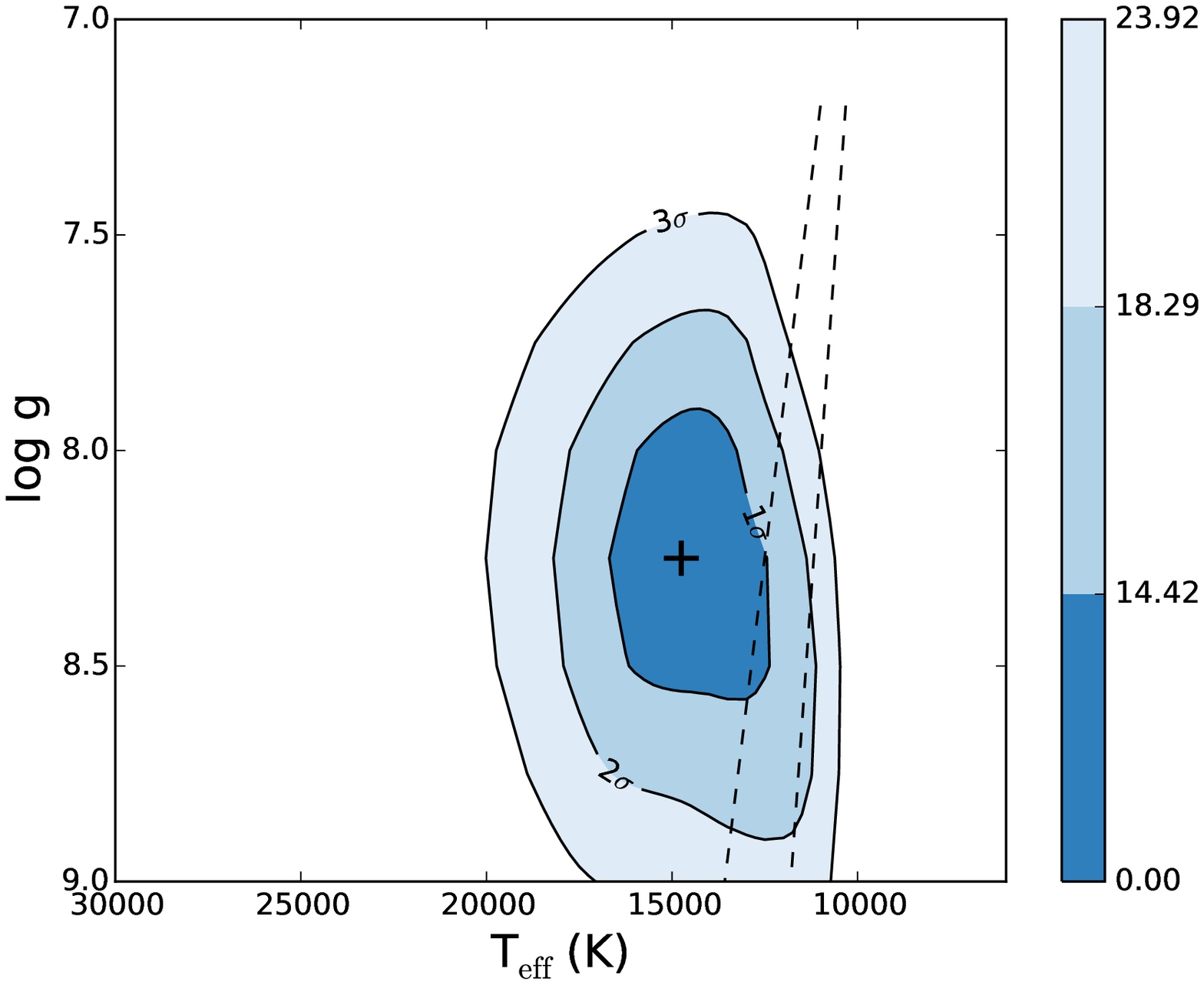}
    \end{subfigure}
    \caption{(a) Stacked Balmer line diagram of the best global model fit for KIC 10777440.  From top to bottom, the corresponding spectral lines are H$\xi$, H$\epsilon$, H$\gamma$, H$\delta$, H$\beta$, and H$\alpha$.  (b) $\chi^2$ contour map for KIC 10777440 of all lines combined in \teff-log $g$ space.  The $+$ indicates the best fit in effective temperature and gravity.  The empirical instability strip is shown by the dashed line.}\label{fig:18486plots}
\end{figure}

KIC 10777440 is a new DA white dwarf in the \emph{Kepler} field.  No previous WD designations or classifications have been found in the literature for this object.  KIC 10777440 has modeled parameters of \teff\ = 14.8 $^{+1.9}_{-2.4}$ kK and log $g$ = 8.25 $^{+0.33}_{-0.35}$ dex.  Figure~\ref{fig:18486plots}a displays the best fit model to each of the Balmer lines.  Figure ~\ref{fig:18486plots}b shows a $\chi^2$ contour map with the 1$\sigma$ to 3$\sigma$ contours; this WD has a moderate probability of being within the instability strip.  

The measured effective temperature of KIC 10777440 does appear to be a little too hot for it to be a pulsator, but it is an instability strip candidate within its 1$\sigma$ errors.  As in the previous case, \emph{Kepler} data is not available to confirm or deny this hypothesis.  \\

\subsubsection{KIC 7879431 (PMI19085+4338)}
KIC 7879431 is a new DA white dwarf in the \emph{Kepler} field.  Best fit parameters of \teff\ = 14.5 $^{+2.2}_{-2.3}$ kK and log $g$ = 8.25 $^{+0.32}_{-0.42}$ dex were determined.  Figure~\ref{fig:19085plots}a displays the best fit model to the Balmer lines of KIC 7879431's optical spectrum.  Figure~\ref{fig:19085plots}b shows the $\chi^2$ contour map in \teff-log $g$ space.  

KIC 7879431 appears more likely to have an effective temperature too hot to be located within the instability strip, however taking into account the 1$\sigma$ errors, this this WD remains a plausible instability strip candidate.  Again, no photometric data is available, so we cannot confirm or deny this hypothesis.

\subsubsection{KIC 9082980 (PMI19179+4524)}
KIC 9082980 is a new DA white dwarf in the \emph{Kepler} field.  Modeled parameters provide \teff\ = 14.0 $^{+1.9}_{-1.9}$ kK and log $g$ = 8.00 $^{+0.42}_{-0.27}$ dex.   Figure~\ref{fig:19179plots}a displays the best fit model to the Balmer lines of the optical spectrum of KIC 9082980.  Figure~\ref{fig:19179plots}b shows the $\chi^2$ contour map in \teff-log $g$ space, where KIC 9082980 falls just outside of the instability strip.  

The modeled temperature and gravity place KIC 9082980 in a region hotter than the blue edge of the instability strip, but the possibility of it being within the strip cannot be ruled out, when the 1$\sigma$ errors are included.  From spectroscopic analysis, KIC 9082980 is an instability strip candidate.  \emph{Kepler} data is not available to confirm or deny this hypothesis.

\subsubsection{KIC 2158770 (PMI19245+3734)}
KIC 2158770 is a known DA white dwarf in the \emph{Kepler} field.  Modeled parameters for KIC 2158770 are, \teff\ = 12.5 $^{+3.3}_{-1.9}$ kK and log $g$ = 8.00 $^{+0.46}_{-0.50}$ dex.  Figure~\ref{fig:19245plots}a shows the best fit model for each of the Balmer lines in the optical spectrum of KIC 2158770.  The $\chi^2$ contour map shown in Fig.~\ref{fig:19245plots}b was used to calculate errors and confidence intervals for the model fitted parameters.  KIC 2158770 is an instability strip candidate because its measurements of effective temperature and gravity, taking the 1$\sigma$ errors into account, can fall within the instability strip.  

A recent study by \citet{kleinman13} used SDSS spectra and measure a \teff\ of 9964 $\pm$ 41 K and a log $g$ of 8.01 $\pm$ 0.054 dex.  This measurement falls at the edge of the $\chi^2$ 3$\sigma$ contour of the modeled parameters of this paper.  SDSS spectra has lower resolution than the Palomar spectra, with $R$ = 1800.  \citet{kleinman13} used a similar set of models and grids, which they fit using an automated process; the fits were checked by eye as well.  Both the fit from this work and the closest fit to the literature are shown in the stacked Balmer line and contour plots in Fig.~\ref{fig:19245plots}.  The best fit model from this paper fits all but H$\beta$ better by eye than the fit from \citet{kleinman13}, which suggests that our estimate is more reliable.  

The periodogram produced from the \emph{Kepler} data does not show any statistically significant ($>$3$\sigma$) periods within the normal range of pulsation periods.  Therefore, this WD is not confirmed to be an instability strip candidate.  However, its proximity to the instability strip potentially makes the star a useful constraint to the strip's blue edge.

\subsubsection{KIC 6672883 (PMI19010+4208)}
KIC 6672883 is a new DA white dwarf in the \emph{Kepler} field with modeled parameters of \teff\ = 16.5$^{+6.6}_{-6.8}$ kK and log $g$ = 7.75$^{+1.14}_{-0.75}$ dex.  Fig.~\ref{fig:19010plots}a shows the best fit model for the fitted Balmer lines.  Fig.~\ref{fig:19010plots}b displays the $\chi^2$ contour map in \teff-log $g$ space, showing large errors in the measurement.  Taking the large 1$\sigma$ error bars on the measurements into account potentially places KIC 6672883 within the instability strip, so it is an instability strip candidate.  

\emph{Kepler} data was available for this target, but did not show any statistically significant periodicity ($>$3$\sigma$) within the period range of pulsations.  We thus confirm KIC 6672883 to be a non-pulsating DA white dwarf. A better spectrum and improved parameters will be necessary to determine if the star can put any useful constraint on the edge of the instability strip.

\subsubsection{KIC 10198116 (PMI19099+4717)}
KIC 10198116 is a known DA white dwarf in the \emph{Kepler} field with best fit modeled parameters of \teff\ = 13.5$^{+3.4}_{-2.1}$ kK and log $g$ = 8.00$^{+0.35}_{-0.31}$ dex.  The nominal temperature appears to be too hot for the star to be in the instability strip, but the errors in the temperature and gravity measurements still makes this a possibility, so we identify KIC 10198116 as an instability strip candidate.Fig.~\ref{fig:19099plots}a shows the best fit model plotted on the Balmer lines.  Fig.~\ref{fig:19099plots}b shows the $\chi^2$ contour map in \teff-log $g$ space of the best fit model and 1, 2, and 3$\sigma$ confidence intervals.  KIC 10198116 has been studied by \citet{ostensen11} and \citet{maoz15}.  The modeled parameters from,  \citet{ostensen11} are \teff\ = 14.2(5) kK and log $g$ = 7.9(3) dex, which is within the 1$\sigma$ errors of the measurements made in this paper.  The comparison of the literature fit and the fit from this work are shown in both plots of Fig.~\ref{fig:19099plots}.  \citet{ostensen11} and \citet{maoz15} do not predict KIC 10198116 to be a pulsator.  

The \emph{Kepler} photometric data does not show any statistically significant periodicity ($>$3$\sigma$) within the normal period range of pulsations.  KIC 10198116 is confirmed to be a non-pulsator, and this may help to put a constraint on the blue edge of the instability strip.

\subsubsection{KIC 11509531 (PMI19320+4925)}
KIC 11509531 is a new DA white dwarf in the \emph{Kepler} field with best fit modeled parameters of \teff\ = 9.25$^{+5.36}_{-1.52}$ kK and log $g$ = 8.25$^{+0.75}_{-1.25}$ dex.  Fig.~\ref{fig:19320plots}a shows the Balmer lines overplotted with the best fit model.  Fig.~\ref{fig:19320plots}b shows the $\chi^2$ contour map in \teff-log $g$ space with confidence intervals.  It can be seen that there are two separate 1$\sigma$ contours, each with its own minimum $\chi^2$ value.  The lowest $\chi^2$ value is actually for parameters \teff\ = 25.0$^{+5.0}_{-8.6}$ kK and log $g$ = 7.00$^{+1.45}_{-0.00}$ dex, but both this and the previously mentioned fits were analyzed by eye and it was determined that the lower effective temperature value provides a better fit to the line cores (comparison shown in both panels of Fig.~\ref{fig:19320plots}).  The $B-V$ and $U-B$ colors are also not blue enough to be consistent with a 25.0 kK sources.  For these reasons, we adopt the lower \teff\ value.  This modeled effective temperature is on the low side for the star to be in the instability strip, but the 1$\sigma$ error makes this a possibility, which leaves KIC 11509531 as an instability strip candidate.  

\emph{Kepler} data was available for this object, but no statistically significant periodicity ($>$3$\sigma$) was found within the normal range of periods for pulsations.  A few peaks were found above 3$\sigma$ significance for longer periods (ones corresponding with variability), but when the data was phased to each of these periods, there was no clear periodic trend.  KIC 11509531 is confirmed to be a non-pulsating DA white dwarf, and one which might place a constraint on the red edge of the instability strip.  Better spectra would be needed again in this case.

\subsubsection{KIC 10213347 (PMI19362+4714)}
KIC 10213347 is a new DA white dwarf in the \emph{Kepler} field.  The modeled parameters for the best fit are \teff\ = 10.3$^{+17.6}_{-2.0}$ kK and log $g$ = 8.00$^{+1.00}_{-1.00}$ dex.  Fig.~\ref{fig:19362plots}a displays the best fit model plotted on the Balmer lines of the observed spectrum.  Fig.~\ref{fig:19362plots}b shows the $\chi^2$ contour map with confidence intervals in \teff-log $g$ space.  The 1$\sigma$ errors are very large, but the modeled effective temperature does lie close to the instability strip and can lie within the instability strip when the errors are taken into account.  We therefore classify KIC 10213347 as an instability strip candidate.  

\emph{Kepler} data, however, does not show any statistically significant periodicity ($>$3$\sigma$) within the range of pulsations.  A single peak is seen with a period corresponding to variability, but when the data is phased to this long period, no plausible periodic trend is apparent.  Therefore, KIC 10213347 is confirmed to be a non-pulsator, but could potentially constrain the red edge of the instability strip.

\subsubsection{KIC 8244398 (PMI19423+4407)} \label{sec:pmi19423}
KIC 8244398 is a new DA white dwarf in the \emph{Kepler} field.  It has modeled parameters of \teff\ = 13.5$^{+7.3}_{-3.5}$ kK and log $g$ = 8.00$^{+0.76}_{-0.90}$ dex.  Fig.~\ref{fig:19423plots}a shows the best fit model overplotted on the Balmer lines of the observed spectrum.  Fig.~\ref{fig:19423plots}b shows the $\chi^2$ contour map in \teff-log $g$ space with 1, 2, and 3$\sigma$ confidence intervals.  The model temperature is on the hot side, but the instability strip is within the 1$\sigma$ error bars, which are relatively large in this case.  When these large errors are taken into account, KIC 8244398 is an instability strip candidate.  

The \emph{Kepler} photometric data produces a periodogram that does not show any possible periods with greater than 3$\sigma$ significance, within the range of pulsations.  We therefore claim KIC 8244398 to be a non-pulsator, but it can help to constrain the blue side of the instability strip.

\subsubsection{KIC 9228724 (PMI19430+4538)}
KIC 9228724 is a new DA white dwarf in the \emph{Kepler} field.  The best fit modeled parameters are \teff\ = 12.8$^{+6.8}_{-2.3}$ kK and log $g$ = 8.25$^{+0.49}_{-0.88}$ dex.  Fig.~\ref{fig:19430plots}a shows the best fit model plotted on the observed spectrum's Balmer lines.  Fig.~\ref{fig:19430plots}b displays the $\chi^2$ contour map in \teff-log $g$ space.  It is apparent that KIC 9228724 lies very close to the blue edge of the instability strip and can lie within the instability strip when the 1$\sigma$ errors are taken into account.  KIC 9228724 is an instability strip candidate.  

The \emph{Kepler} data do not show any statistically significant periodicity ($>$3$\sigma$) within the likely range of periods for pulsating DA white dwarfs.  KIC 9228724 is confirmed to be a non-pulsator.  Again this star may place a constraint on the blue edge of the instability strip.

\subsection{Non-Members of the Instability Strip} \label{sec:nonpuls}
\subsubsection{KIC 10649118 (PMI18553+4755)}
KIC 10649118 is a new DA white dwarf in the \emph{Kepler} field.  Modeled parameters produced \teff\ = 8.50 $^{+0.82}_{-0.69}$ kK and log $g$ = 8.00 $^{+0.64}_{-0.77}$ dex.  Figure~\ref{fig:18553plots}a shows the best fit model to each of the Balmer lines.  Figure~\ref{fig:18553plots}b displays the 1-3$\sigma$ contours of the best fit parameters for log $g$ and \teff.  

The $\chi^2$ contour map in \teff-log $g$ space shows two 1$\sigma$ minima.  The other minimum has \teff\ = 27.0 $^{+3.0}_{-5.5}$ kK and log $g$ = 7.00 $^{+0.57}_{-0.00}$ dex.  Both minima were examined by eye and the minimum with \teff\ = 9.00 kK was visually determined to have the better overall fit of the line cores, even though \teff\ = 27.0 kK had the lower $\chi^2$ value.  $B-V$ and $U-B$ colors also suggest an effective temperature consistent with 8.50 kK.  The two fits can be seen on the Balmer line and contour plots in Fig.~\ref{fig:18553plots}.  

Neither of the above measurements fall within the instability strip (shown in Fig.~\ref{fig:18553plots}b), therefore this WD is not an instability strip candidate.  There is no \emph{Kepler} photometry available to confirm or deny this hypothesis.

\subsubsection{KIC 4242459 (PMI19002+3922)}
KIC 4242459 is confirmed to be a DA white dwarf and modeled parameters of \teff\ = 9.50 $^{+0.22}_{-0.13}$ kK and log $g$ = 8.25 $^{+0.09}_{-0.05}$ dex were determined.  KIC 4242459 was studied in the optical/IR by \citet{zuckerman03}, who measured \teff\ = 9470 K with a fixed value of log $g$ = 8.00 using photometry, which closely matches our measurements; it is within our 1$\sigma$ error contour.  Figure~\ref{fig:19002plots}a shows the best fit model from this work and the closest match to the best fit model from the literature.  Figure~\ref{fig:19002plots}b shows the $\chi^2$ contour map of KIC 4242459 in the \teff-log $g$ plane with confidence intervals.  

Modeled parameters for KIC 4242459 place it below the red edge of the instability strip, even when the 1$\sigma$ errors are taken into consideration.  This WD is not an instability strip candidate and photometric data confirms that there is no statistically significant variability ($>$3$\sigma$) on pulsation timescales.

\subsubsection{KIC 11604781 (PMI19141+4936)}\label{sec:pmi19141}
KIC 11604781 is a known DA white dwarf with a companion of unknown spectral type.  Our best fit modeled parameters are \teff\ = 9.50 $^{+0.55}_{-0.26}$ K and log $g$ = 8.50 $^{+0.32}_{-0.10}$ dex.  \citet{ostensen11} found this object to be a DA5 with \teff\ = 9.1(5) kK and log $g$ = 8.3(3), which is within 3$\sigma$ of our result.  Figure~\ref{fig:19141plots}a shows the best fit model to the spectrum of KIC 11604781.  Figure~\ref{fig:19141plots}b shows the $\chi^2$ contour map in \teff-log $g$ space, with 1, 2 and 3$\sigma$ contours marked.  The fits from this work and the closest match to the fit of \citet{ostensen11} are compared in both Figs.~\ref{fig:19141plots}a and \ref{fig:19141plots}b.  

\begin{figure}[h]
\centering
\includegraphics[scale=0.35]{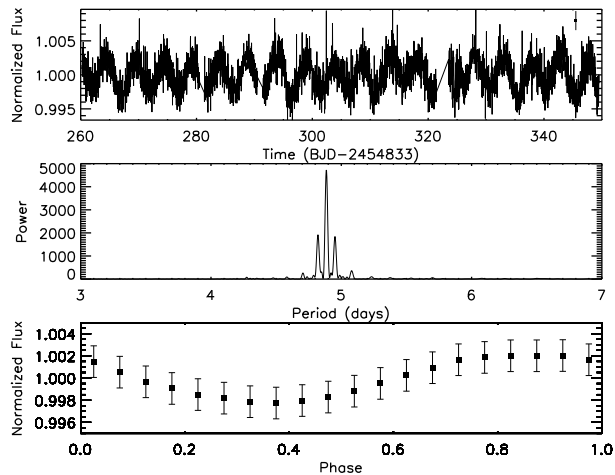}
\caption{Long cadence section of the \emph{Kepler} light curve, followed by the Lomb-Scargle periodogram, which shows the most likely period of KIC 11604781.  A typical error bar is shown in the upper right hand corner of the upper plot.  The bottom plot is the binned LC data phased with a period of 4.89 days.  The periodogram was produced from all the available \emph{Kepler} data (5 quarters) and the phased curve contains $\sim$500 points in each bin.}  
\label{fig:19141lsp}
\end{figure}

From its modeled parameters, KIC 11604781 is not an instability strip candidate, however we do detect variability in its light curve.  A period of 4.89 days is extracted from the photometric data, which is well outside of the normal period range for pulsations.  Figure~\ref{fig:19141lsp} shows a section of the \emph{Kepler} light curve, the Lomb-Scargle periodogram of the entire light curve, and the phased and binned light curve for KIC 11604781.  
A representative point with a typical error bar is located in the top right corner of the light curve.  KIC 11604781 has been studied by a few others \citep{ostensen11,maoz15}, who found the same periodicity (4.89$\pm$0.02 days) without a concrete explanation as to its cause.  Explanations of possible companions and the causation of this period are provided in \citet{maoz15}.  The light curve variation is most likely an orbital variation and is determined to not be due to pulsations.  No statistically significant, greater than 3$\sigma$, periodicity was found within the range of DA white dwarf pulsations.

\subsubsection{KIC 8682822 (PMI19173+4452)}
KIC 8682822 is a known DA white dwarf and was studied by both \citet{ostensen10} and \citet{maoz15}.  The best fit modeled parameters from this work yield \teff\ = 19.5 $^{+1.3}_{-1.1}$ kK and log $g$ = 8.75 $^{+0.14}_{-0.10}$ dex.  \citet{ostensen10} studied this WD as a compact pulsator candidate.  \citet{ostensen10} measures a \teff\ = 23.1 kK and log $g$ = 8.5 dex, which is hotter than our modeled effective temperature, and is just outside of the 3$\sigma$ contour of our result.  Figure~\ref{fig:19173plots}a shows the best fit model of \teff\ and log $g$ for the Balmer lines in the optical spectrum of KIC 8682822.  Figure~\ref{fig:19173plots}b shows the $\chi^2$ contour map of KIC 8682822 in \teff-log $g$ space; it falls outside of the instability strip.  The fits from this work and the closest match to the \citet{ostensen10} fit are compared in both Figs.~\ref{fig:19173plots}a and \ref{fig:19173plots}b.  

No statistically significant period, greater than 3$\sigma$ significance, is found in the \emph{Kepler} data, which agrees with KIC 8682822's modeled effective temperature being too hot to fall within the instability strip.  \citet{maoz15} measures a mass of $\sim$0.8 - 1.2\msun, with extremely small amplitude variations leading to a period of 4.7$\pm$0.3d.  Photometric data confirms that there are no variations on the order of pulsations, and we do not recover the 4.7 day period that \citet{maoz15} claims.  KIC 8682822 is confirmed to be a non-pulsating DA white dwarf.

\subsubsection{KIC 7129927 (PMI19409+4240)} \label{sec:pmi19409}
KIC 7129927 is a known DA white dwarf.  We determine \teff\ = 9.50 $^{+0.45}_{-0.31}$ kK and log $g$ = 8.25 $^{+0.30}_{-0.15}$ dex.  \citet{ostensen11} estimates \teff\ and log $g$ from observations of two separate years; \teff\ = 23314 $\pm$ 212 and log $g$ = 7.280 $\pm$ 0.050 (year 1) and \teff\ = 24191 $\pm$ 332 and log $g$ = 7.120 $\pm$ 0.060 (year 2).  \citet{ostensen11} claims this object is a composite DA2+DA3 white dwarf binary system.  Figure~\ref{fig:19409plots}a shows the best fit model to the Balmer lines of KIC 7129927 and includes the closest fit to both of the \citet{ostensen11} fits for comparison.  Figure~\ref{fig:19409plots}b shows the $\chi^2$ contour map in \teff-log $g$ space of KIC 7129927, which falls outside of the instability strip.  The fits from \citet{ostensen11} are noted on the contour map.  

Our measurements do not match well with those of \citet{ostensen11}.  Balmer line best fit diagrams were examined for the closest models in our grid to the \teff\ and log $g$ from \citet{ostensen11}; \teff\ = 23.0 kK, log $g$ = 7.25 dex and \teff\ = 24.0 kK, log $g$ = 7.00 dex.  The best fit we obtained, fits the cores much better.  It is apparent though, from Fig.~\ref{fig:19409plots}b that there is a 2$\sigma$ and 3$\sigma$ contour around a central \teff\ = 21.5 kK and log $g$ = 7.33 dex.  The measurements from \citet{ostensen11} fall into the secondary 3$\sigma$ contour of our $\chi^2$ plot.  $B-V$ and $U-B$ colors suggest an effective temperature consistent with the measurements from this work.  The discrepancy in the model fits here could result from the fact that KIC 7129927 might be a composite system and our single WD model fit is not appropriate.  

The modeled temperature estimate from this work places KIC 7129927 just below the red edge of the instability strip, but this WD is not an instability strip candidate.  It does not coincide with the instability strip when the 1$\sigma$ errors are considered.  \emph{Kepler} data confirms there is no detectable, statistically significant ($>$3$\sigma$) variability within the range of pulsations in the photometric data, noted by \citet{maoz15} as well.

\subsubsection{KIC 6042560 (Blue18)}
KIC 6042560 is a new DA white dwarf in the \emph{Kepler} field.  The modeled parameters for this WD are \teff\ = 6.75$^{+0.76}_{-0.75}$ kK and log $g$ = 7.75$^{+1.11}_{-0.75}$ dex.  Fig.~\ref{fig:19010plots}a displays the best fit model overplotted on the five measured Balmer lines.  Fig.~\ref{fig:19010plots}b shows the contour $\chi^2$ map in \teff-log $g$ space with the best fit marked.  It is clear from this figure that KIC 6042560 is far too cool to be located within the instability strip, therefore it is not an instability strip candidate.  

Photometric data from \emph{Kepler} shows no statistically significant periodicity ($>$3$\sigma$) within the range of pulsations.  KIC 6042560 is confirmed to be a non-pulsating DA white dwarf.

\subsubsection{KIC 7346018 (Blue1903)}
KIC 7346018 is a new DA white dwarf in the \emph{Kepler} field.  The best fit modeled parameters are \teff\ = 30.0$^{+0.0}_{-3.5}$ kK and log $g$ = 8.50 $^{+0.50}_{-0.82}$ dex.  It is noted here that the positive error of 0.0 kK is not a true error, but the upper limit of the model grid..  The effective temperature measurement may thus be underestimated, as 30.0 kK is the absolute upper bound of the DA white dwarf models.  Fig.~\ref{fig:blue1903plots}a displays the best model fit overplotted on the observed spectral Balmer lines.  Fig.~\ref{fig:blue1903plots}b shows the $\chi^2$ contour map with the 1, 2 , and 3$\sigma$ confidence intervals in \teff-log $g$ space.  It is clear that KIC 7346018 lies far from the instability strip in effective temperature and it is not an instability strip candidate.  We do note, however, that there is a possibility this white dwarf is a hot pulsating DA white dwarf (or DAV), a new class of DAVs classified by \citet{kurtz13}, which have temperatures around 30 kK.  

The \emph{Kepler} data available for KIC 7346017 however does not show any periodicity with greater than 3$\sigma$ significance within the period range of pulsations.  This DA white dwarf is confirmed as a non-pulsator.

\subsubsection{KIC 8612751 (PMI19060+4446)}
KIC 8612751 is a new DA white dwarf in the \emph{Kepler} field.  The best fit model provided measurements of \teff\ = 7.50$^{+1.35}_{-1.08}$ kK and log $g$ = 8.00$^{+1.00}_{-1.00}$ dex.  These parameters place KIC 8612751 below the instability strip in temperature, and even taking errors into account, it is not an instability strip candidate.  Fig.~\ref{fig:19060plots}a shows the best fit model plotted on the observed spectrum for all of the Balmer lines.  Fig.~\ref{fig:19060plots}b shows the $\chi^2$ contour map with the best fit marked in \teff-log $g$ space.  A second set of 1, 2, and 3$\sigma$ contours can be seen at the upper left corner of the $\chi^2$ plot.  Both minima of each 1$\sigma$ contour were examined by eye and the fit of \teff\ = 30.0 kK and log $g$ = 7.00 dex was determined to not be as good of a fit as the previously noted values (see Fig.~\ref{fig:19060plots} for comparison).  $B-V$ and $U-B$ colors also provide an effective temperature which agrees with \teff\ = 7.50 kK.  

\emph{Kepler} data are available for this white dwarf, but no periodic modulation with greater than 3$\sigma$ significance was found from the period analysis.  KIC 8612751 is confirmed to be a non-pulsating DA white dwarf.

\subsubsection{KIC 4829241 (HSH08)} \label{sec:hsh08}
KIC 4829241 is a known DA white dwarf in the \emph{Kepler} field with modeled parameters of \teff\ = 19.5$^{+1.5}_{-1.6}$ kK and log $g$ = 8.00$^{+0.29}_{-0.24}$ dex.  Several studies of KIC 4829241 have been completed and two of these have measured effective temperatures and gravities.  \citet{ostensen11} measured an effective temperature of 19.4(5) kK and gravity of 7.8(3) dex.  \citet{zhao13} found a modeled effective temperature of 20376$\pm$345 K and gravity of 7.93$\pm$0.06 dex.  Both of the previously determined values are within the 1$\sigma$ errors of the measurements from this paper. Fig.~\ref{fig:hsh08aplots}a shows the best fit model plotted on the observed spectrum's Balmer lines.  Fig.~\ref{fig:hsh08aplots}b displays the contour $\chi^2$ map in \teff-log $g$ space with $1 - 3\sigma$ confidence intervals.  Both plots in Fig.~\ref{fig:hsh08aplots} show the comparison of the closest model to the literature fits with the fit from this work.  As model fits for this work and the fit from \citet{ostensen11} are the essentially same, the alternative fit represents the model from \citet{zhao13}.    KIC 4829241 has a temperature too hot to be within the instability strip and is therefore not an instability strip candidate.  

\begin{figure*}[h]
\centering
\includegraphics[scale=0.35]{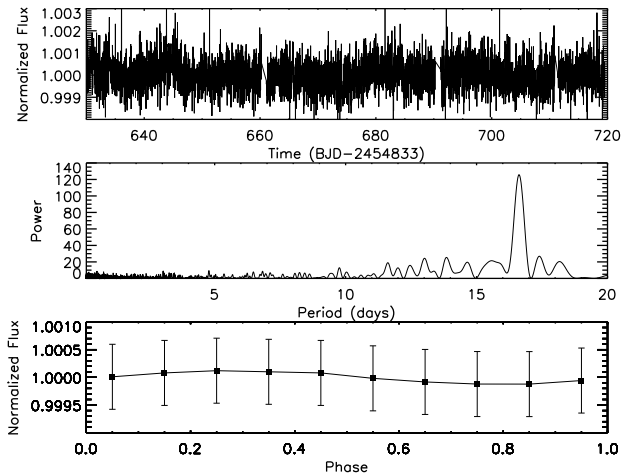}
\caption{Long cadence section of the \emph{Kepler} light curve for KIC 4829241, followed by the Lomb-Scargle periodogram, which show the most likely period in the pulsation range.  The bottom plot is the binned LC data phased with a period of 16.61 days.  The periodogram was produced from all the available \emph{Kepler} data (13 quarters) and the phased curve contains $\sim$500 points in each bin.}  
\label{fig:HSH08alsp}
\end{figure*}

\emph{Kepler} data show no significant periods, greater than 3$\sigma$ significance, within the range of pulsation periods.  A long period variation of 16.61 days was found, which is well above 3$\sigma$ significance (Fig.~\ref{fig:HSH08alsp}).  However, taking the errors into account, the data could very well have no variation.  Neither \citet{ostensen11} nor \citet{zhao13} found a pulsation period for this white dwarf.  KIC 4829241 is confirmed to be a non-pulsating DA white dwarf.  

\subsubsection{KIC 6212123 (HSH24)}
KIC 6212123 is a new DA white dwarf in the \emph{Kepler} field.  Best fit modeled parameters yield \teff\ = 6.25$^{+0.59}_{-0.25}$ kK and log $g$ = 8.50$^{+0.50}_{-0.82}$ dex.  Fig.~\ref{fig:hsh24plots}a displays the best fit model overplotted on the observed spectrum of the measured Balmer lines.  Fig.~\ref{fig:hsh24plots}b shows the contour $\chi^2$ map in \teff-log $g$ space with confidence intervals.  The measured effective temperature of KIC 6212123 is too cool to fall within the instability strip, even taking the 1$\sigma$ errors into account.  KIC 6212123 is not an instability strip candidate.  

\emph{Kepler} data shows no statistically significant periodicity ($>$3$\sigma$) in the range of pulsations for this white dwarf.  KIC 6212123 is confirmed to be a non-pulsating DA white dwarf.

\subsubsection{KIC 3354599 (HSH32)}
KIC 3354599 is a new DA white dwarf in the \emph{Kepler} field with modeled parameters of \teff\ = 6.00$^{+0.92}_{-0.00}$ kK and log $g$ = 8.50$^{+0.50}_{-1.09}$ dex.  Again, the 0.0 kK error is due to the limit of the model grid.  Fig.~\ref{fig:hsh32plots}a shows the best fit model plotted on top of the observed spectrum of all the measured Balmer lines (H$\alpha$ to H$\xi$).  Fig.~\ref{fig:hsh32plots}b shows the $\chi^2$ contour map in \teff-log $g$ space with the best fit model and confidence intervals marked.  The modeled effective temperature of KIC 3354599 is too cool to be within the instability strip, even taking the 1$\sigma$ errors into account, therefore it is not an instability strip candidate.  

Available \emph{Kepler} photometric data does not show any statistically significant, greater than 3$\sigma$, periodicity within the normal range of pulsations.  KIC 3354599 is confirmed to be a non-pulsating DA white dwarf.

\subsubsection{KIC 10149875 (HSH36)}
KIC 10149875 is a new DA white dwarf in the \emph{Kepler} field.  The best fit modeled parameters are \teff\ = 30.0$^{+0.0}_{-0.8}$ kK and log $g$ = 9.00$^{+0.00}_{-0.29}$ dex.  Both the 0.0 kK and 0.00 dex errors mark the upper bounds of the models fit to these DA white dwarfs.  Fig.~\ref{fig:hsh36plots}a shows the best fit model plotted on the observed spectrum's measured Balmer lines.  Fig.~\ref{fig:hsh36plots}b shows the $\chi^2$ contour map in \teff-log $g$ space with confidence intervals.  The minimum $\chi^2$ values within each 1$\sigma$ contour were examined by eye and the best fit was determined to be the fit with \teff\ = 30.0 kK, which also had the lowest $\chi^2$ value.  The alternative $\chi^2$ minimum fit was \teff\ = 7.25$^{+0.49}_{-0.62}$ kK and log $g$ = 9.00$^{+0.00}_{-0.62}$ dex.  $B-V$ and $U-B$ colors are most consistent with the lower effective temperature value of 9.00 kK.  The comparison of the two fits is shown in both plots in Fig.~\ref{fig:hsh36plots}.  The effective temperature of KIC 10149875 is too hot to be within the instability strip, therefore it is not an instability strip candidate.  This white dwarf is also a possible hot DAV \citep{kurtz13}.  

\emph{Kepler} photometric data shows no statistically significant ($>$3$\sigma$) periods within the range of pulsations.  KIC 10149875 is confirmed to be a non-pulsating DA white dwarf.

\section{Conclusion}\label{sec:conclusion}
We presented modeled effective temperatures and gravities for 23 DA white dwarfs in the \emph{Kepler} field using ground-based spectroscopic observations.  Seven of the twenty three WDs had been previously studied and are here confirmed to be DA white dwarfs.  The remaining 16 are newly classified DA white dwarfs in the \emph{Kepler} field.  

Eighteen of the total twenty three DA white dwarfs were supplemented by \emph{Kepler} photometric data, making it possible to search for modulations from white dwarf pulsation modes.  Eleven WDs are found from spectroscopic measurements to be instability strip candidates when their 1$\sigma$ errors are taken into account and twelve were determined to be non-pulsators (see Fig.~\ref{fig:instastrip}).  Out of the 11 instability strip candidates, 7 had photometric data and none of these were seen to have statistically significant periods within the normal period range of DA white dwarf pulsations.  Of the 12 non-pulsators, 10 had photometric data from \emph{Kepler} and were determined to be photometrically stable from both spectroscopic and photometric analysis.  These 10 photometrically stable DA white dwarfs can be used as photometric calibrators in the \emph{Kepler} field.  One more non-pulsating DA white dwarf, KIC 11604781, had \emph{Kepler} data, but has a companion, so it is not photometrically stable.  We do not confirm any pulsators in this sample of 23 DA white dwarfs.  

\section*{Acknowledgments}
White dwarf models were provided by Dr. Detlev Koester at the University of Kiel in Germany.  Balmer/Lyman lines in the models were calculated with the modified Stark broadening profiles of \citet{tremberg09}, kindly made available by the authors.  Some of the data presented in this paper were obtained from the Mikulski Archive for Space Telescopes (MAST). STScI is operated by the Association of Universities for Research in Astronomy, Inc., under NASA contract NAS5-26555. Support for MAST for non-HST data is provided by the NASA Office of Space Science via grant NNX13AC07G and by other grants and contracts.  This paper includes data collected by the Kepler mission. Funding for the Kepler mission is provided by the NASA Science Mission directorate.  This research has made use of the NASA Exoplanet Archive, which is operated by the California Institute of Technology, under contract with the National Aeronautics and Space Administration under the Exoplanet Exploration Program.  This research is based on observations at Kitt Peak National Observatory, National Optical Astronomy Observatory, which is operated by the Association of Universities for Research in Astronomy (AURA) under cooperative agreement with the National Science Foundation.  This paper made use of the facilities at the Kitt Peak National Observatory and the Palomar Observatory.  The authors would like to thank the telescope operators and staff at both of these institutions.  We thank Mark Everett for help with the Mayall observations and Jay Holberg and David Sing for help with the Bok observations.  

\bibliographystyle{mnras}
\bibliography{DAWD_Paper_revised}

\newpage

\appendix
\section{Additional Figures}
{\color{white}Words.} \\

\begin{figure}[H]
    \begin{subfigure}[]{}
        \includegraphics[width=0.46\textwidth]{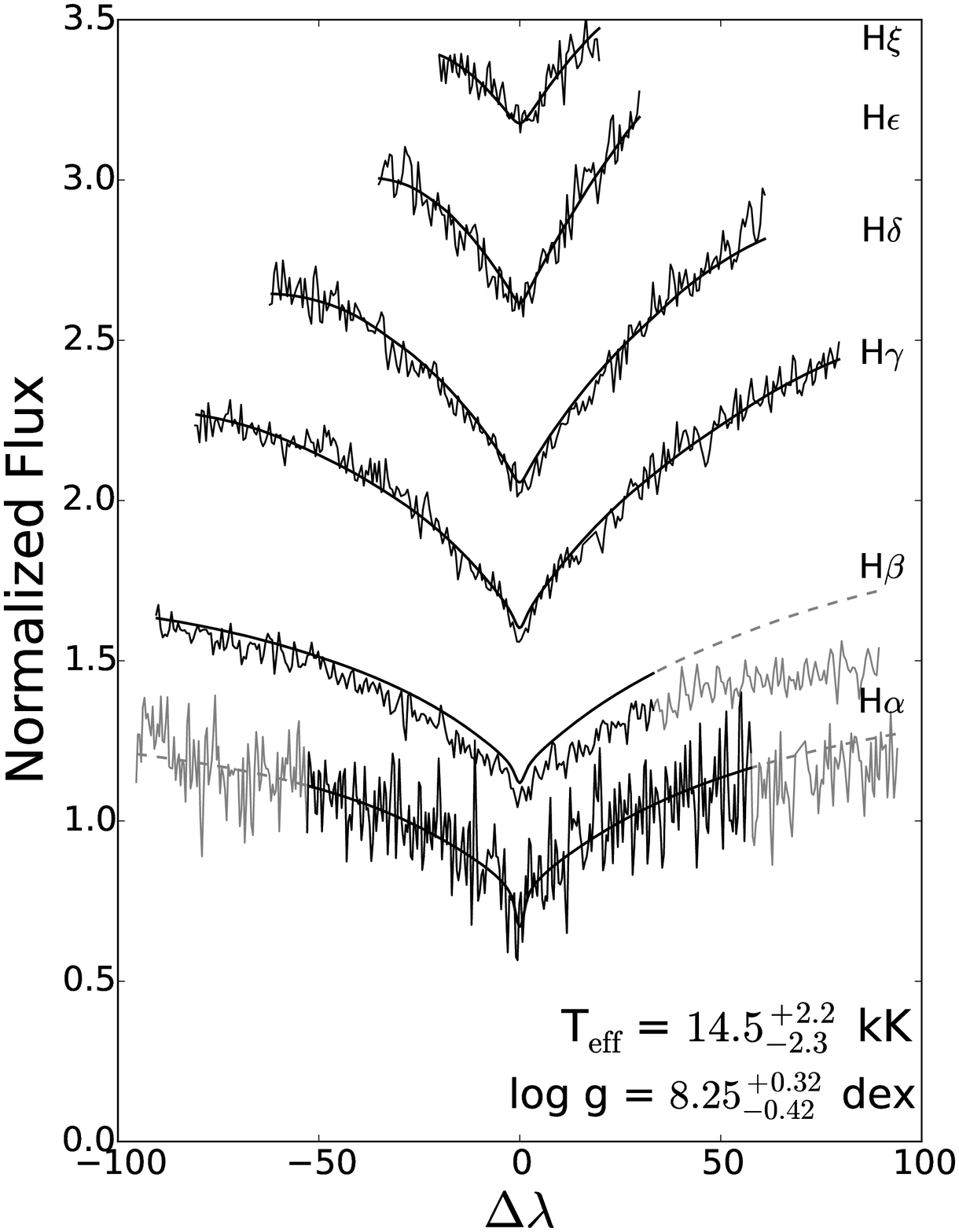}
    \end{subfigure}
    ~  
    \begin{subfigure}[]{}
        \includegraphics[width=0.47\textwidth]{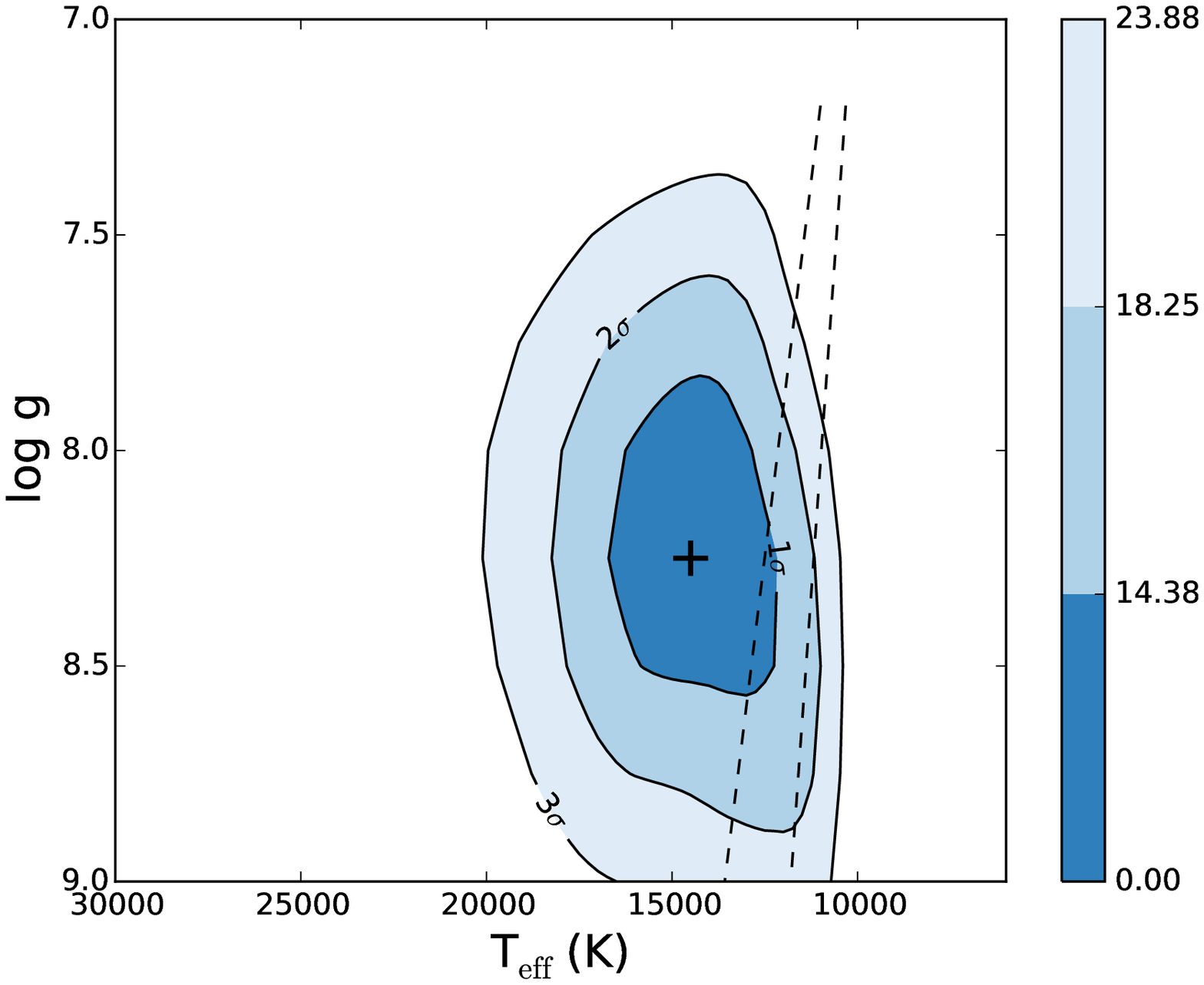}
    \end{subfigure}
    \caption{Same as Fig.~\ref{fig:18486plots}, but for KIC 7879431.}\label{fig:19085plots}
\end{figure}

\begin{figure}
    \begin{subfigure}[]{}
        \includegraphics[width=0.46\textwidth]{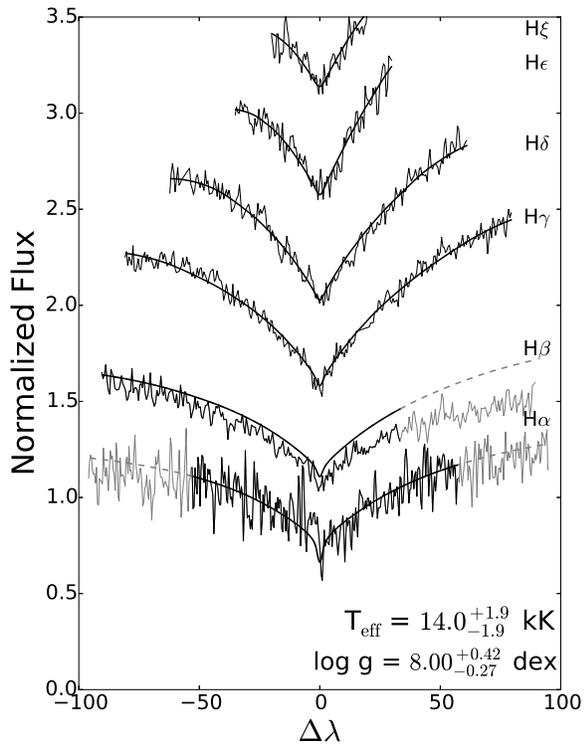}
    \end{subfigure}
    ~  
    \begin{subfigure}[]{}
        \includegraphics[width=0.47\textwidth]{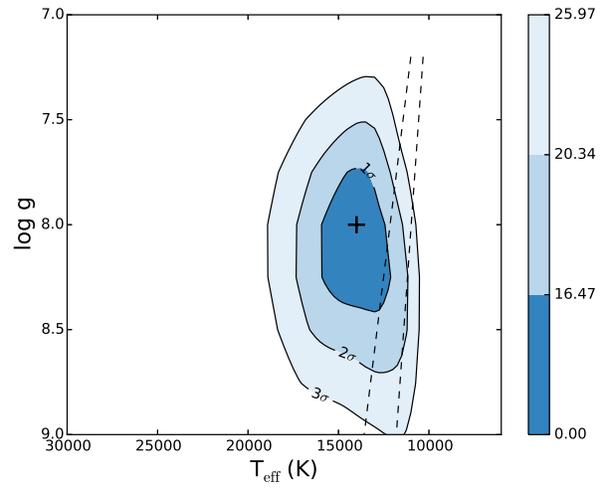}
    \end{subfigure}
    \caption{Same as Fig.~\ref{fig:18486plots}, but for KIC 9082980.}\label{fig:19179plots}
\end{figure}

\begin{figure}
    \begin{subfigure}[]{}
        \includegraphics[width=0.46\textwidth]{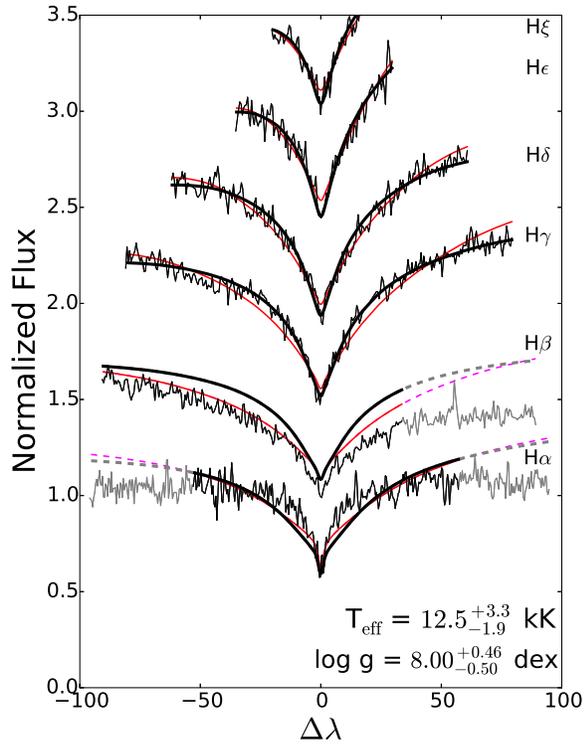}
    \end{subfigure}
    ~  
    \begin{subfigure}[]{}
        \includegraphics[width=0.47\textwidth]{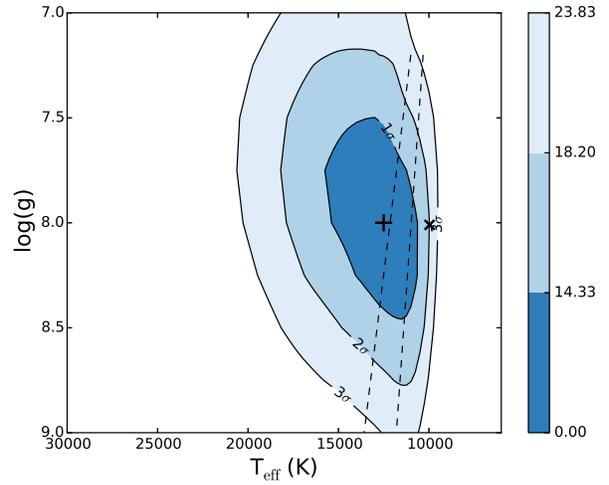}
    \end{subfigure}
    \caption{(a) Stacked Balmer line diagram of the best global model fit (thick black line) for KIC 2158770.  The red line fit corresponds to the modeled parameters from \citet{kleinman13}.  (b) $\chi^2$ contour map for KIC 2158770 of all lines combined in \teff-log $g$ space.  Again, the $+$ corresponds to the best fit from this work and the error contour to show overlap with the instability strip.  The $\times$ shows the modeled parameters from \citet{kleinman13}.}\label{fig:19245plots}
\end{figure}

\begin{figure}
    \begin{subfigure}[]{}
        \includegraphics[width=0.46\textwidth]{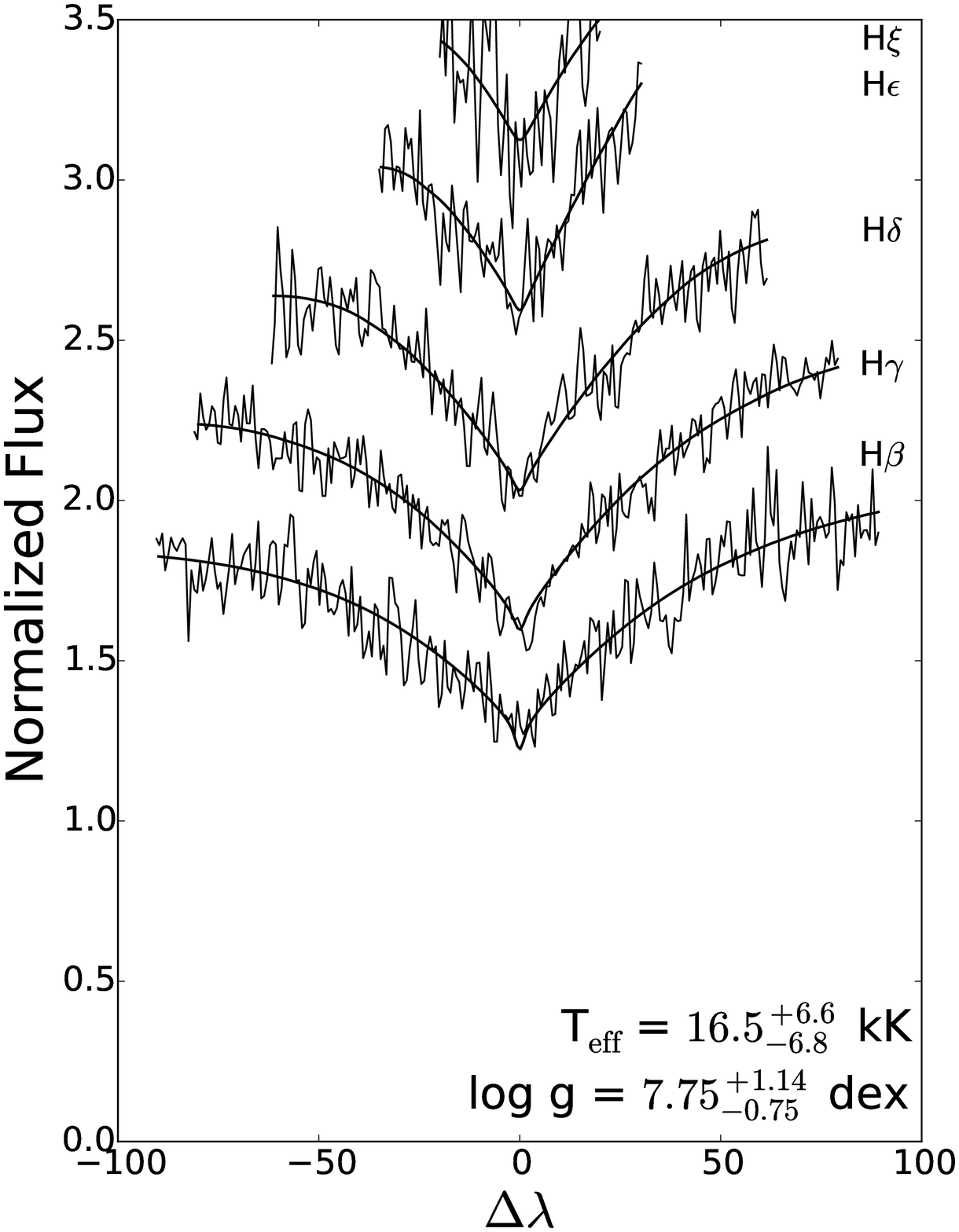}
    \end{subfigure}
    ~  
    \begin{subfigure}[]{}
        \includegraphics[width=0.47\textwidth]{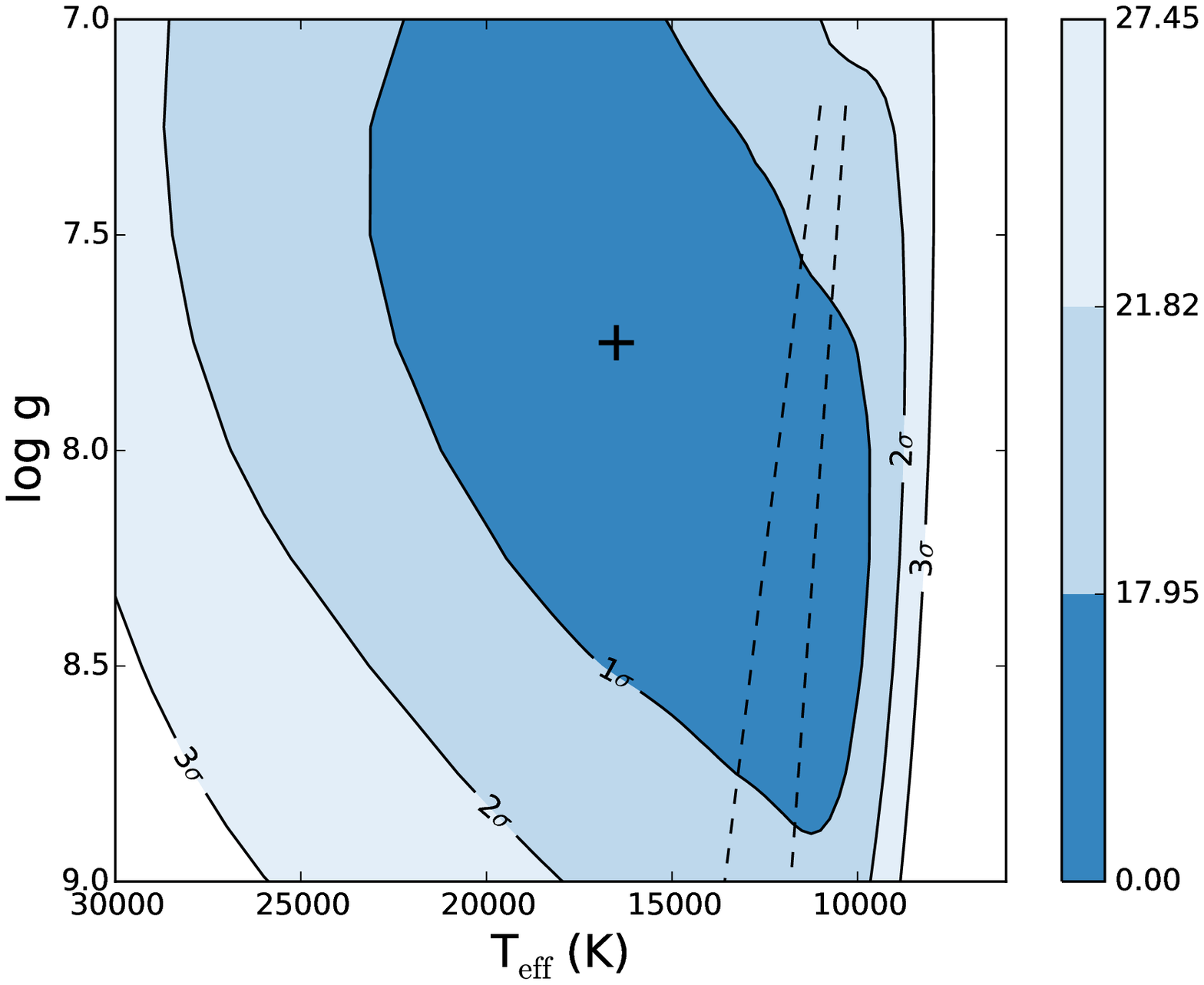}
    \end{subfigure}
    \caption{(a) Stacked Balmer line diagram of the best global model fit for KIC 6672883.  From top to bottom the corresponding spectral lines are H$\xi$, H$\epsilon$, H$\gamma$, H$\delta$ and H$\beta$.  (b) $\chi^2$ contour map for KIC 6672883 of all lines combined in \teff-log $g$ space.  The $+$ indicates the best fit in effective temperature and gravity.  The empirical instability strip is shown by the dashed line.}\label{fig:19010plots}
\end{figure}

\begin{figure}
    \begin{subfigure}[]{}
        \includegraphics[width=0.46\textwidth]{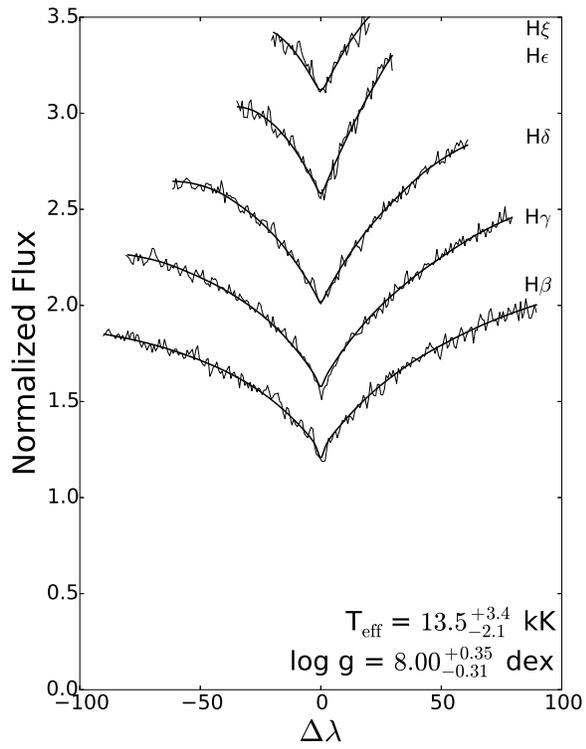}
    \end{subfigure}
    ~  
    \begin{subfigure}[]{}
        \includegraphics[width=0.47\textwidth]{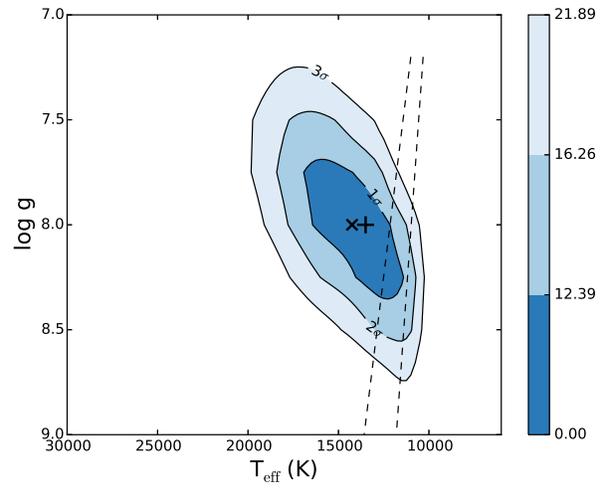}
    \end{subfigure}
    \caption{Same as Fig.~\ref{fig:19010plots}, but for KIC 10198116.  The red line fit in (a) and the $\times$ in (b) show the modeled parameters from \citet{ostensen11}.  }\label{fig:19099plots}
\end{figure}

\begin{figure}
    \begin{subfigure}[]{}
        \includegraphics[width=0.46\textwidth]{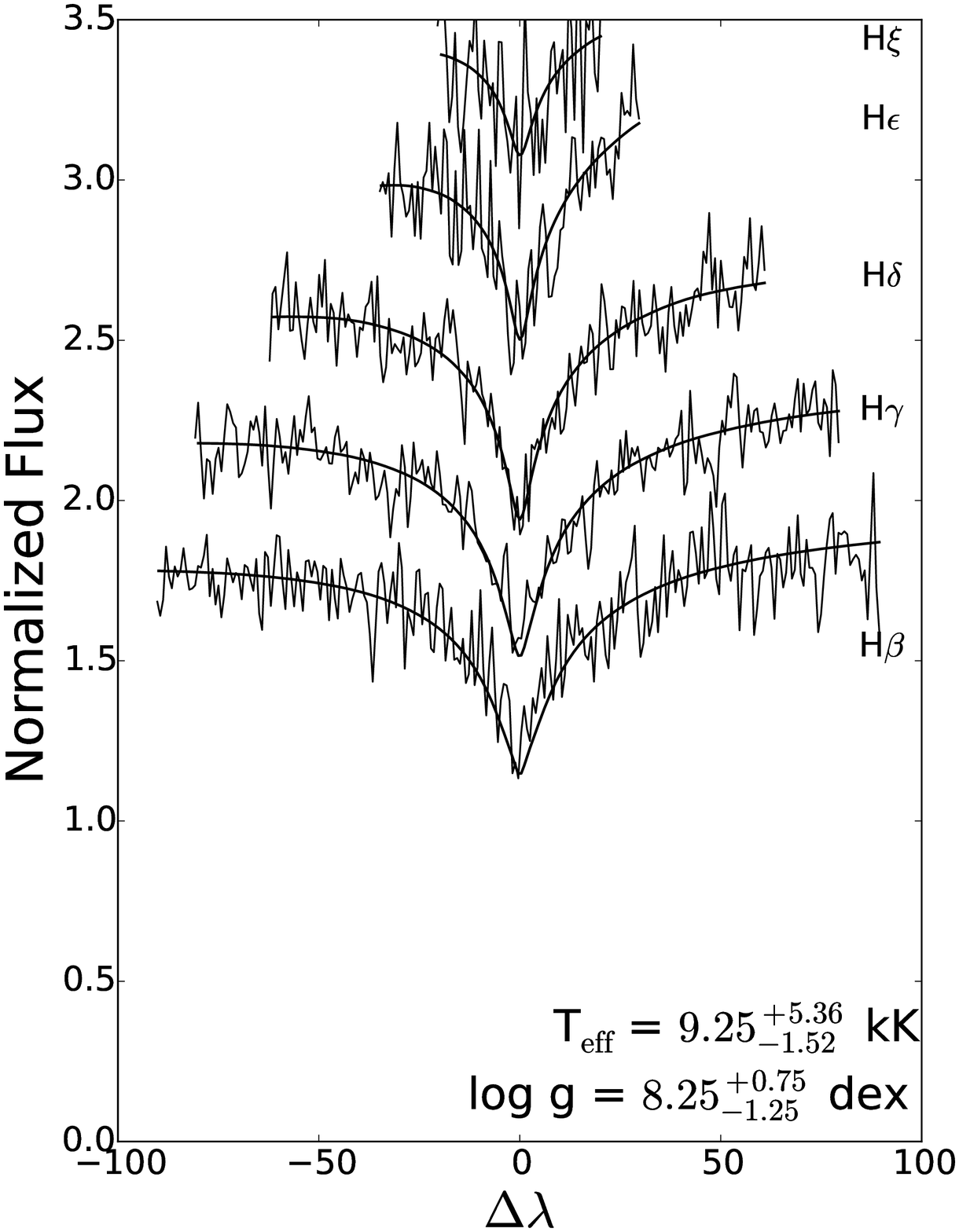}
    \end{subfigure}
    ~  
    \begin{subfigure}[]{}
        \includegraphics[width=0.47\textwidth]{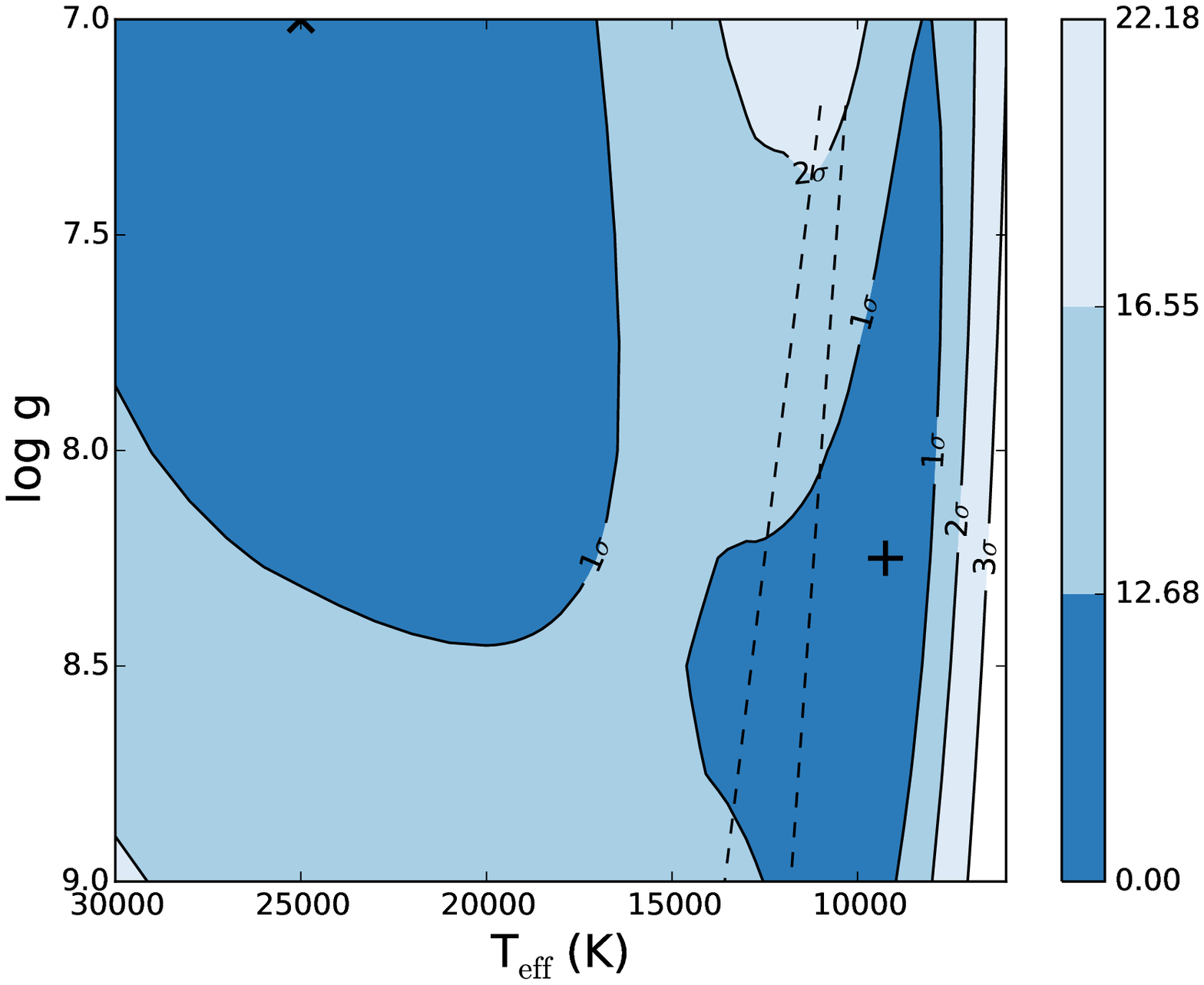}
    \end{subfigure}
    \caption{Same as Fig.~\ref{fig:19010plots}, but for KIC 11509531.  The red line fit in (a) and the $\times$ in (b) correspond to the other $\chi^2$ minimum of \teff\ = 25.0 kK and log $g$ = 7.00 dex.  }\label{fig:19320plots}
\end{figure}

\begin{figure}
    \begin{subfigure}[]{}
        \includegraphics[width=0.46\textwidth]{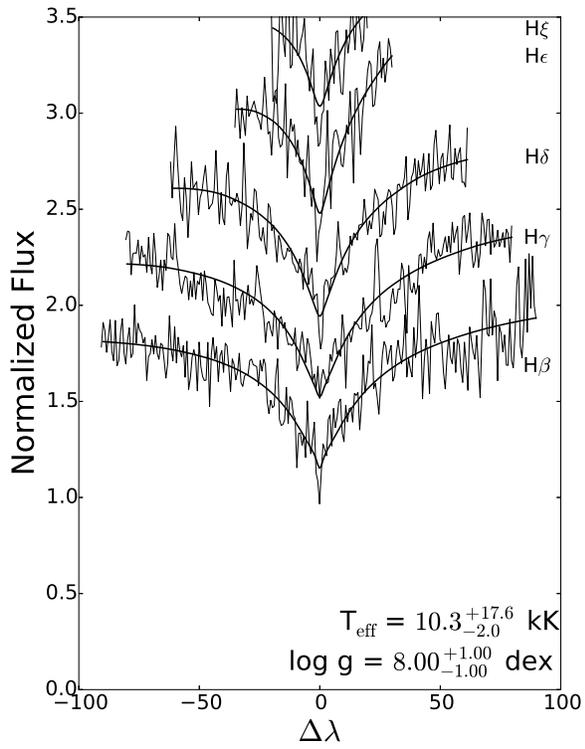}
    \end{subfigure}
    ~  
    \begin{subfigure}[]{}
        \includegraphics[width=0.47\textwidth]{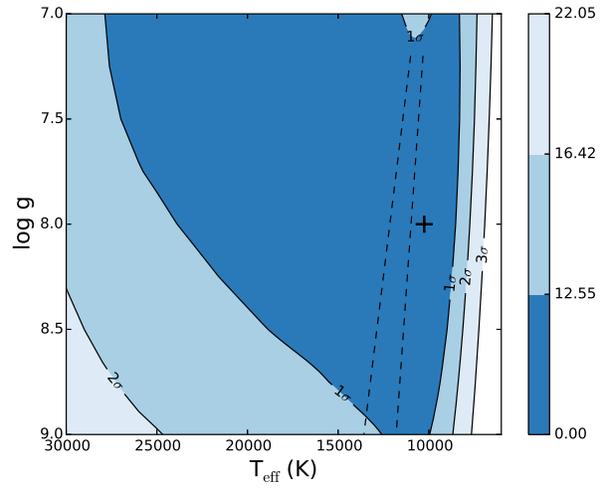}
    \end{subfigure}
    \caption{Same as Fig.~\ref{fig:19010plots}, but for KIC 10213347.}\label{fig:19362plots}
\end{figure}

\begin{figure}
    \begin{subfigure}[]{}
        \includegraphics[width=0.46\textwidth]{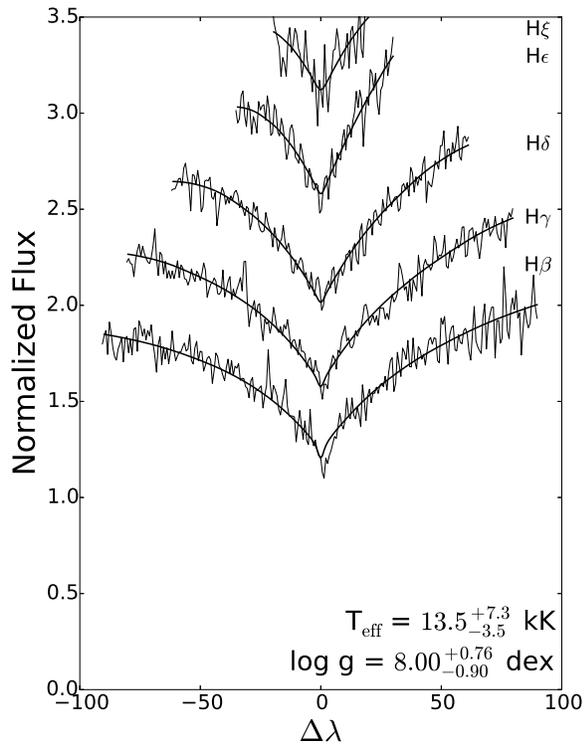}
    \end{subfigure}
    ~  
    \begin{subfigure}[]{}
        \includegraphics[width=0.47\textwidth]{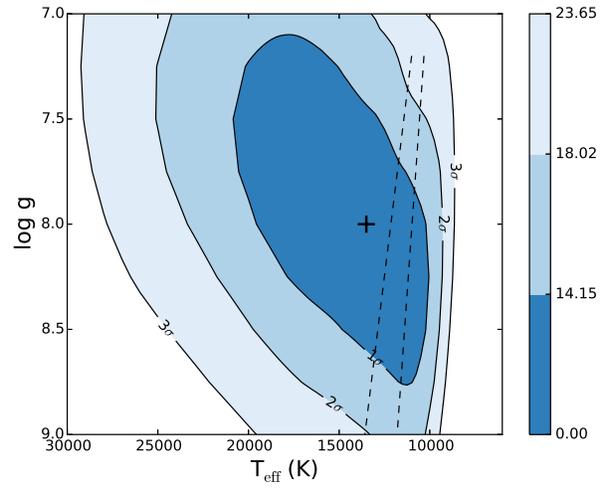}
    \end{subfigure}
    \caption{Same as Fig.~\ref{fig:19010plots}, but for KIC 8244398.}\label{fig:19423plots}
\end{figure}

\begin{figure}
    \begin{subfigure}[]{}
        \includegraphics[width=0.46\textwidth]{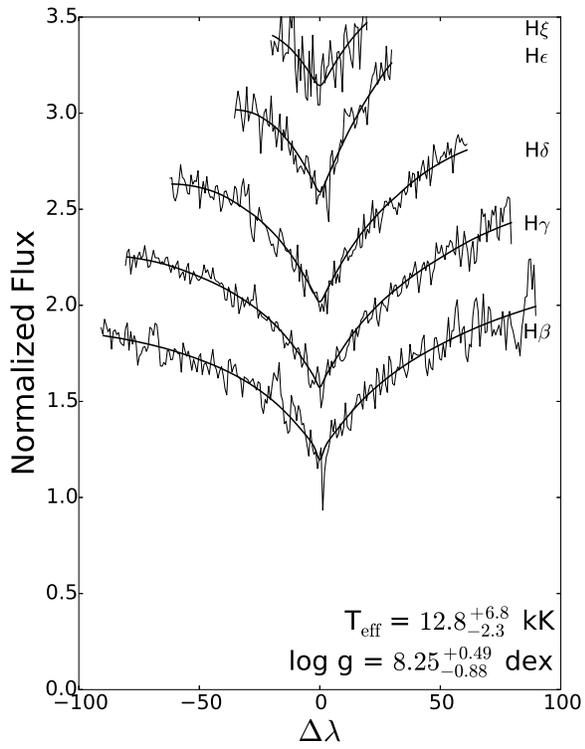}
    \end{subfigure}
    ~  
    \begin{subfigure}[]{}
        \includegraphics[width=0.47\textwidth]{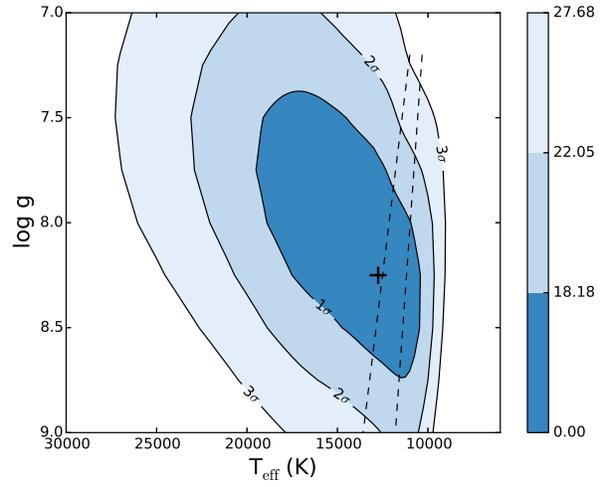}
    \end{subfigure}
    \caption{Same as Fig.~\ref{fig:19010plots}, but for KIC 9228724.}\label{fig:19430plots}
\end{figure}

\begin{figure}
    \begin{subfigure}[]{}
        \includegraphics[width=0.46\textwidth]{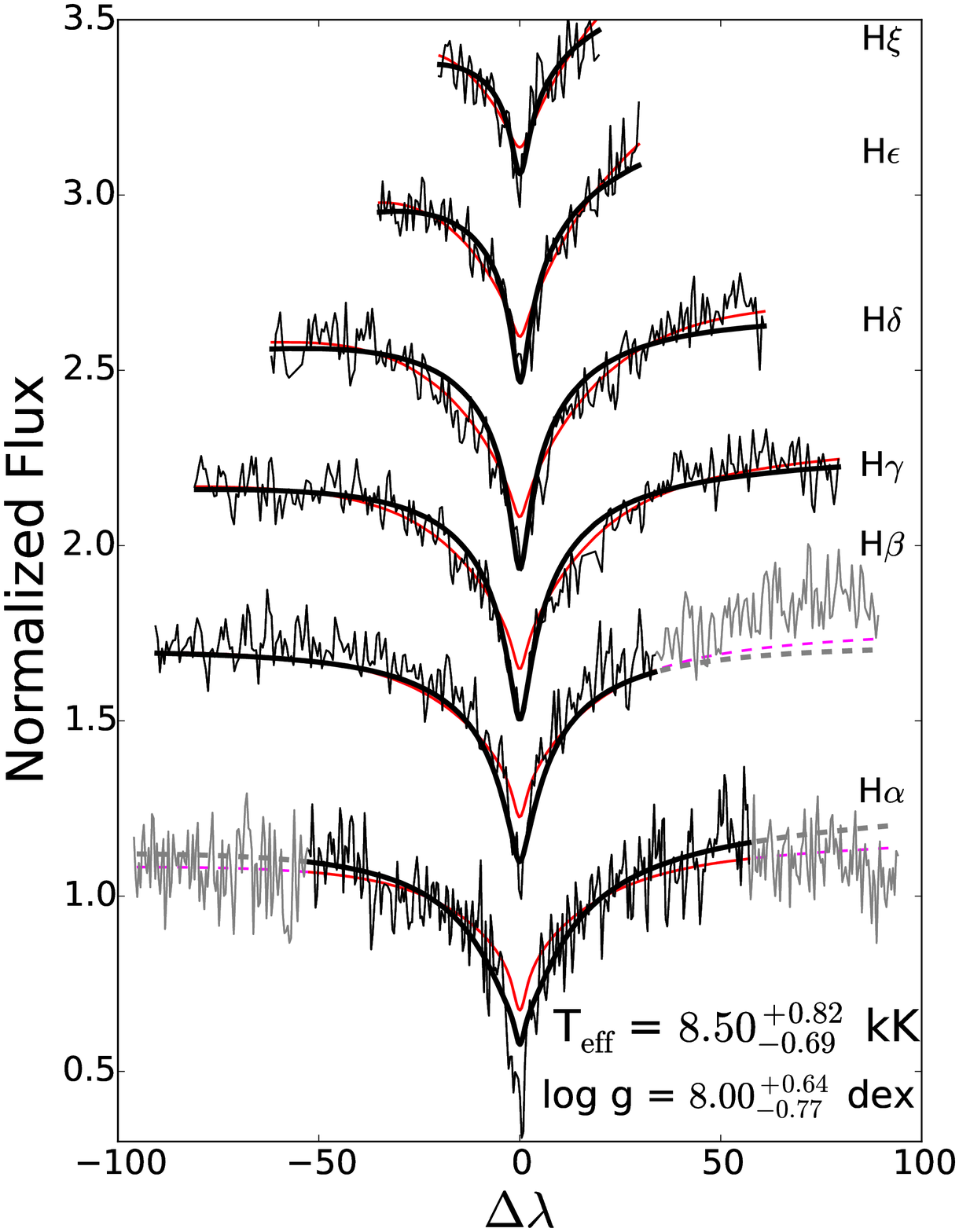}
    \end{subfigure}
    ~  
    \begin{subfigure}[]{}
        \includegraphics[width=0.47\textwidth]{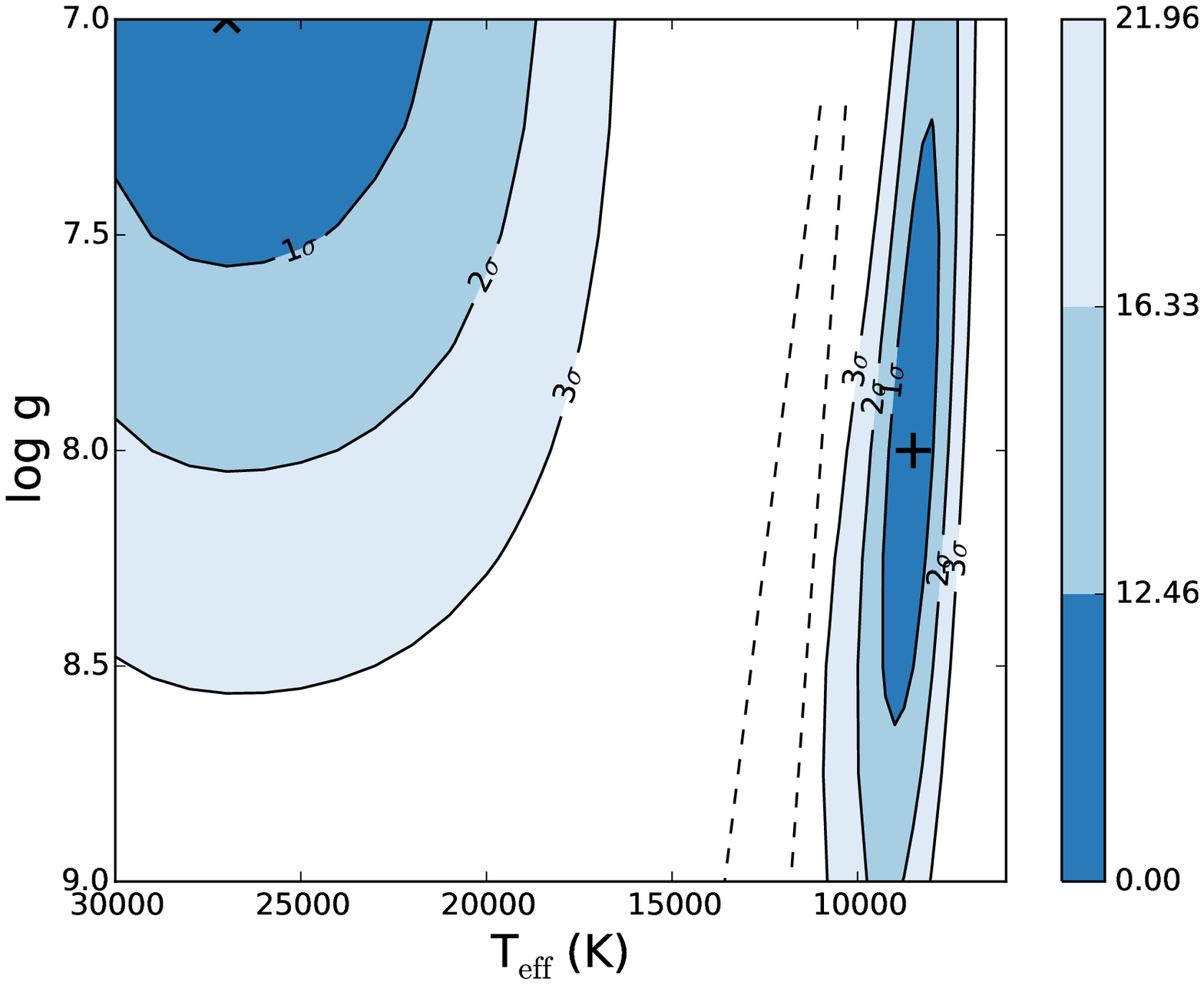}
    \end{subfigure}
    \caption{Same as Fig.~\ref{fig:18486plots}, but for KIC 10649118.  The red line fit in (a) and the $\times$ in (b) corresponds to the secondary $\chi^2$ minimum of \teff\ = 27.0 kK and log $g$ = 7.00 dex.}\label{fig:18553plots}
\end{figure}

\begin{figure}
    \begin{subfigure}[]{}
        \includegraphics[width=0.46\textwidth]{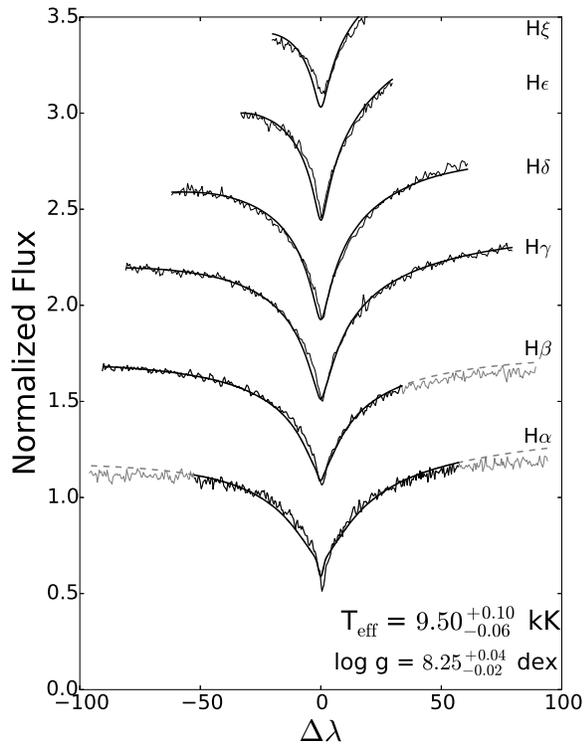}
    \end{subfigure}
    ~  
    \begin{subfigure}[]{}
        \includegraphics[width=0.47\textwidth]{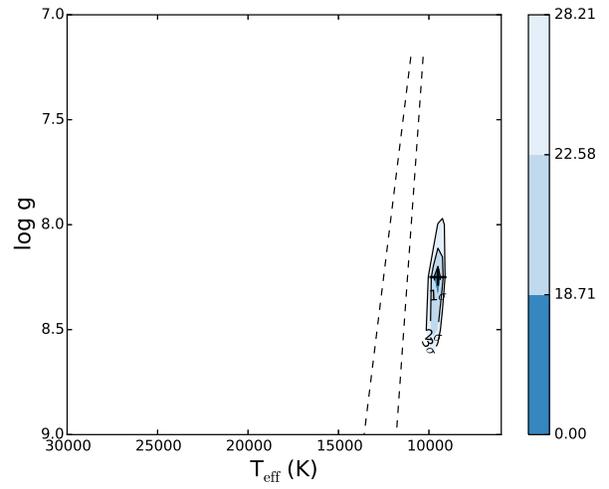}
    \end{subfigure}
    \caption{Same as Fig.~\ref{fig:18486plots}, but for KIC 4242459.}\label{fig:19002plots}
\end{figure}

\begin{figure}
    \begin{subfigure}[]{}
        \includegraphics[width=0.46\textwidth]{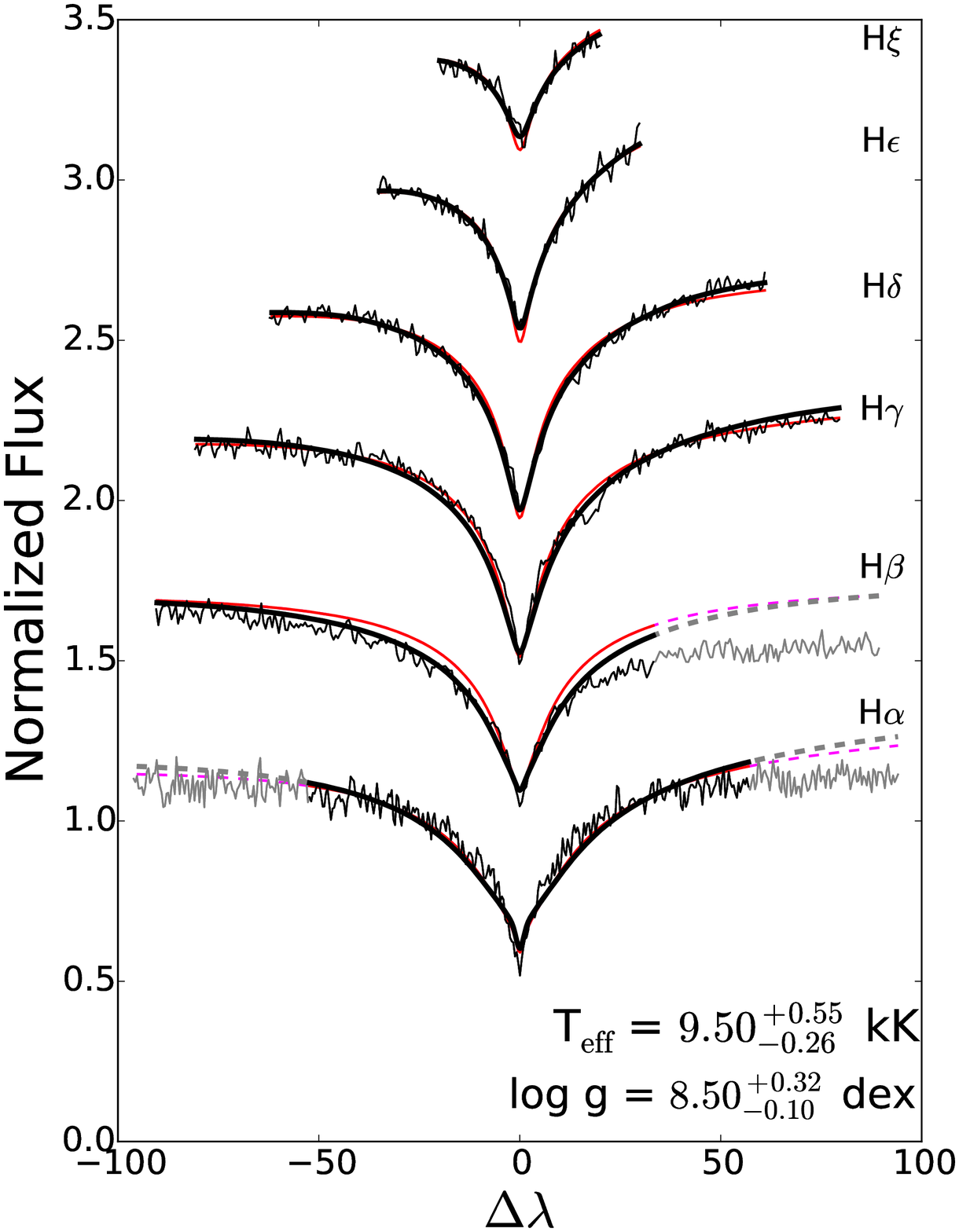}
    \end{subfigure}
    ~  
    \begin{subfigure}[]{}
        \includegraphics[width=0.47\textwidth]{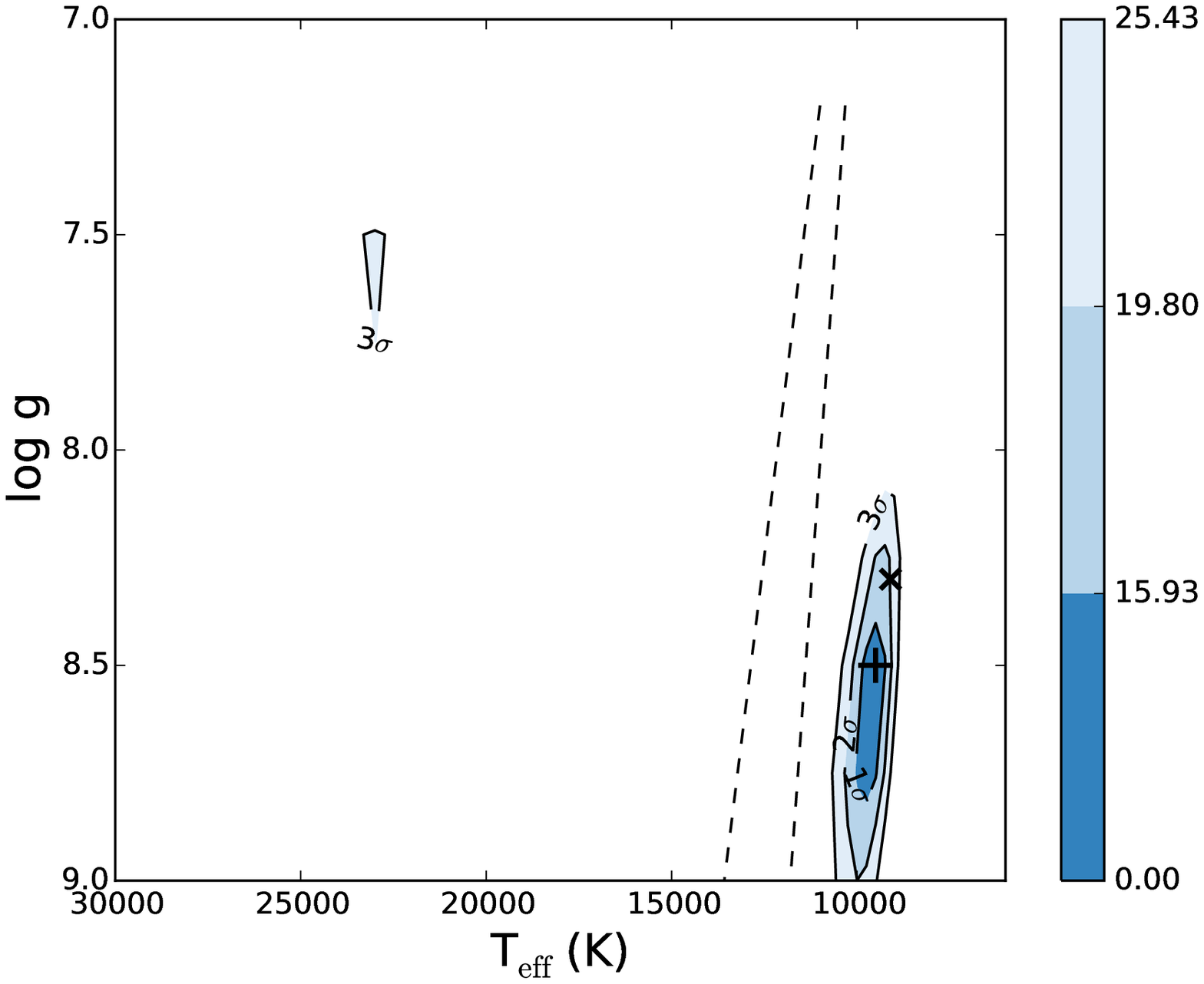}
    \end{subfigure}
    \caption{Same as Fig.~\ref{fig:18486plots}, but for KIC 11604781.  The red line fit in (a) and the $\times$ in (b) shows the modeled parameters from \citet{ostensen11}.}\label{fig:19141plots}
\end{figure}

\begin{figure}
    \begin{subfigure}[]{}
        \includegraphics[width=0.46\textwidth]{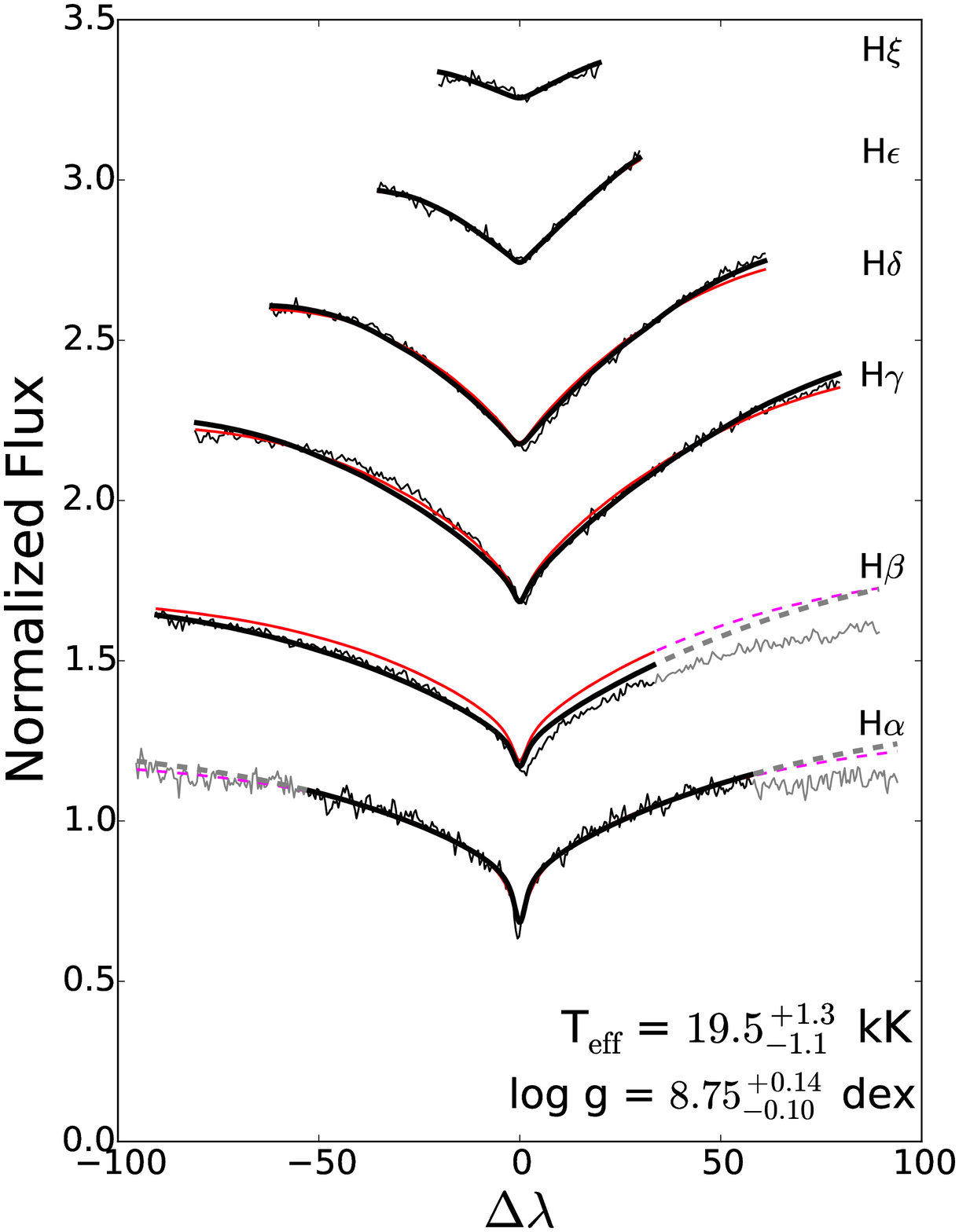}
    \end{subfigure}
    ~  
    \begin{subfigure}[]{}
        \includegraphics[width=0.47\textwidth]{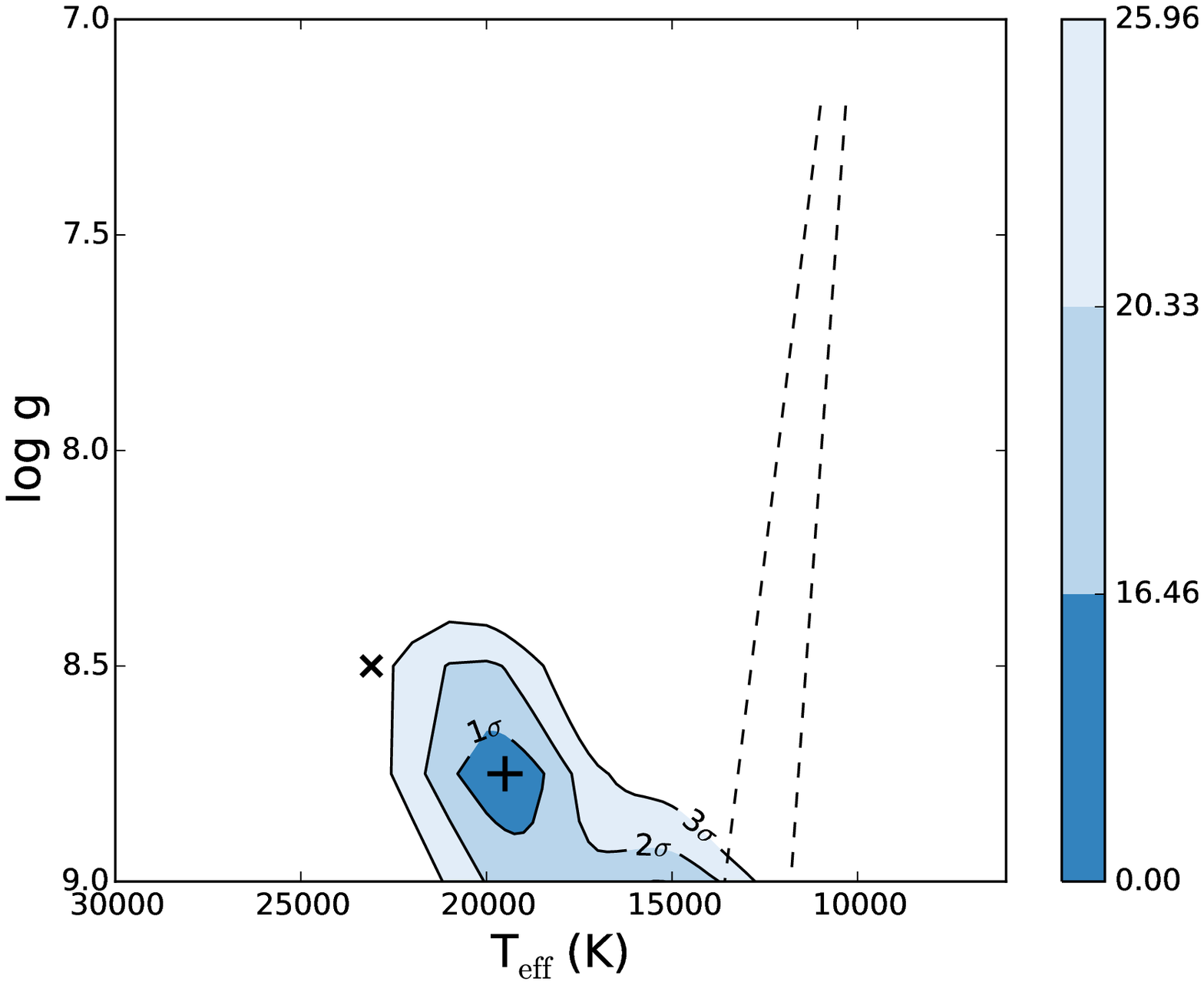}
    \end{subfigure}
    \caption{Same as Fig.~\ref{fig:18486plots}, but for KIC 8682822.  The red line fit in (a) and the $\times$ in (b) shows the modeled parameters from \citet{ostensen10}.}\label{fig:19173plots}
\end{figure}

\begin{figure}
    \begin{subfigure}[]{}
        \includegraphics[width=0.46\textwidth]{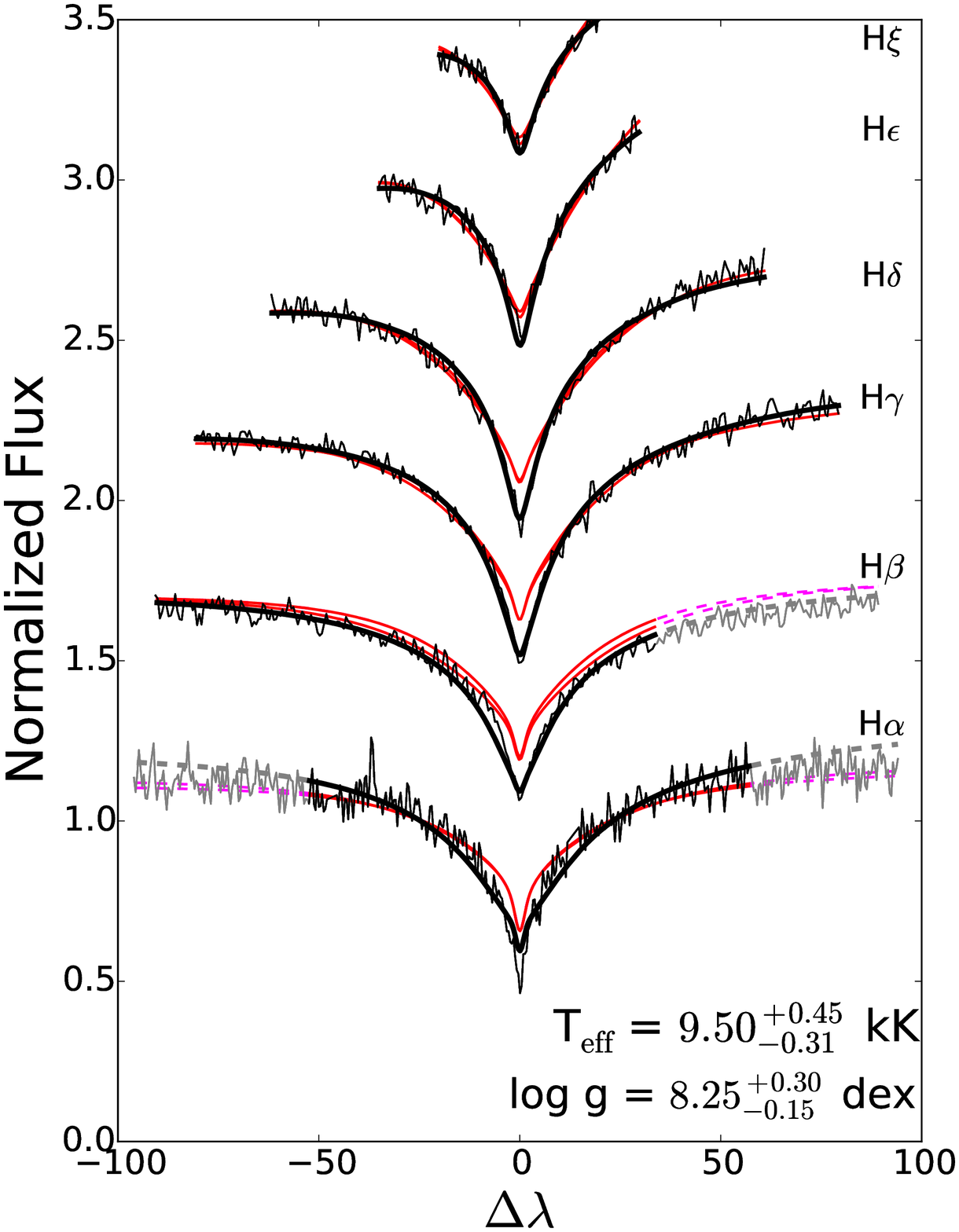}
    \end{subfigure}
    ~  
    \begin{subfigure}[]{}
        \includegraphics[width=0.47\textwidth]{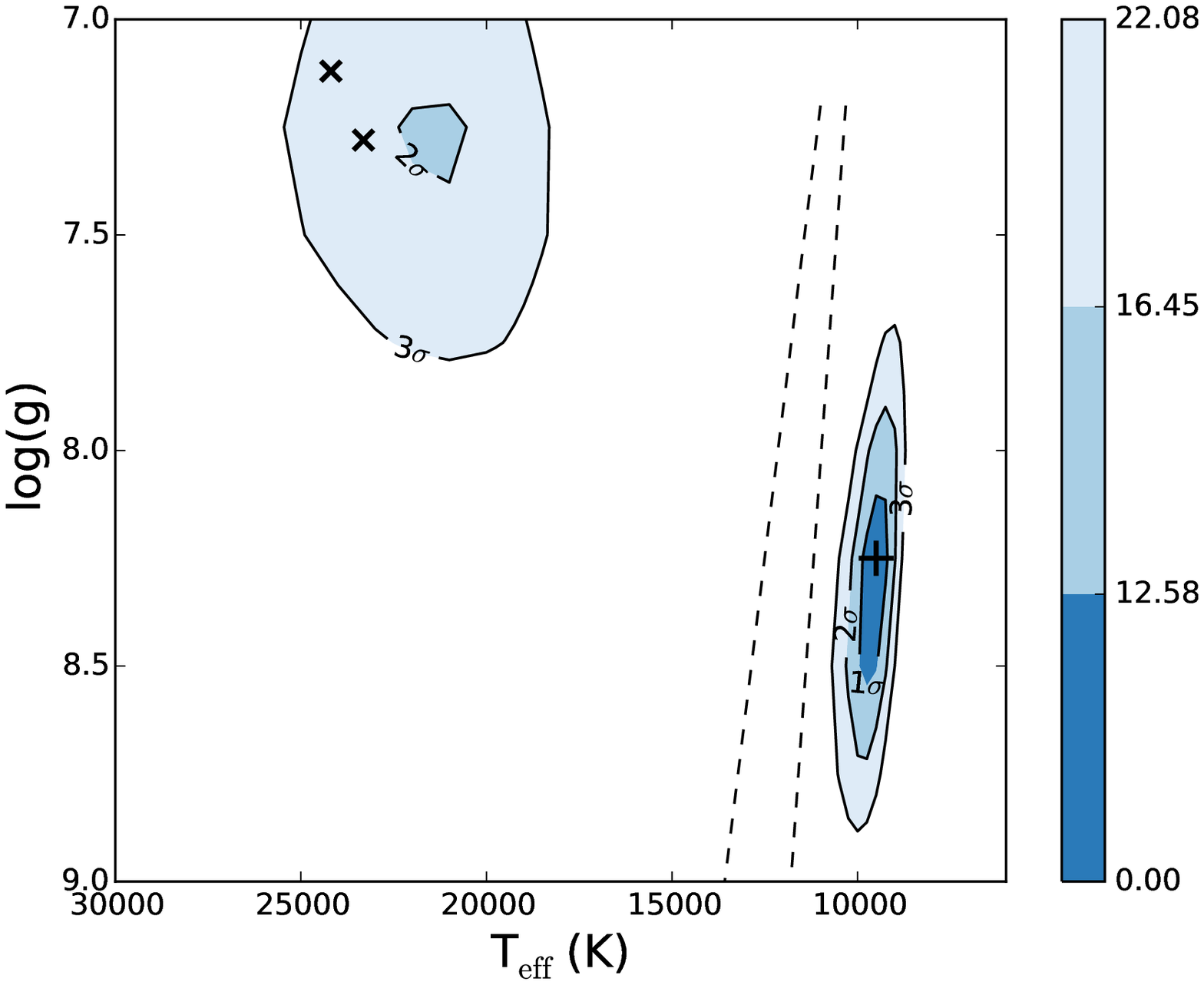}
    \end{subfigure}
    \caption{Same as Fig.~\ref{fig:18486plots}, but for KIC 7129927.  The two red line fits in (a) and the $\times$'s in (b) show the modeled parameters from \citet{ostensen11}.}\label{fig:19409plots}
\end{figure}

\begin{figure}
    \begin{subfigure}[]{}
        \includegraphics[width=0.46\textwidth]{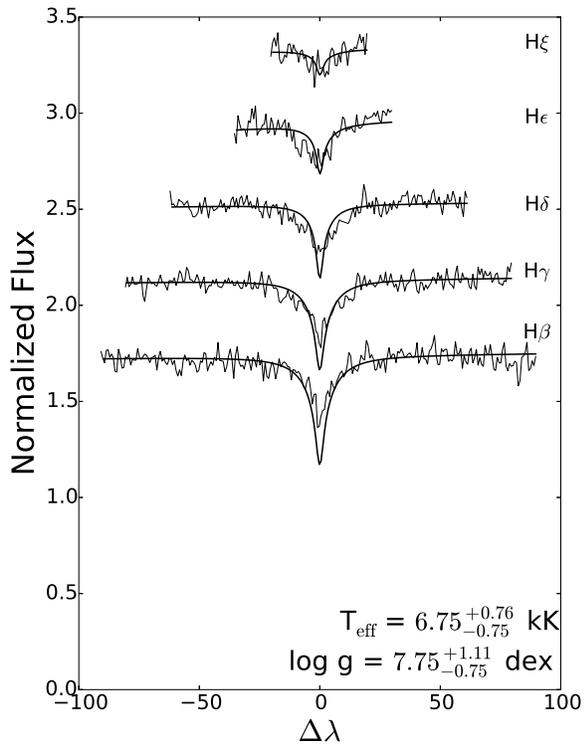}
    \end{subfigure}
    ~  
    \begin{subfigure}[]{}
        \includegraphics[width=0.47\textwidth]{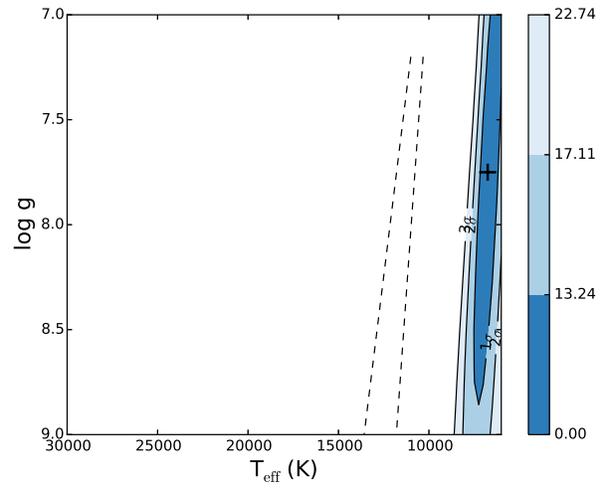}
    \end{subfigure}
    \caption{Same as Fig.~\ref{fig:19010plots}, but for KIC 6042560.}\label{fig:blue18plots}
\end{figure}

\begin{figure}
    \begin{subfigure}[]{}
        \includegraphics[width=0.46\textwidth]{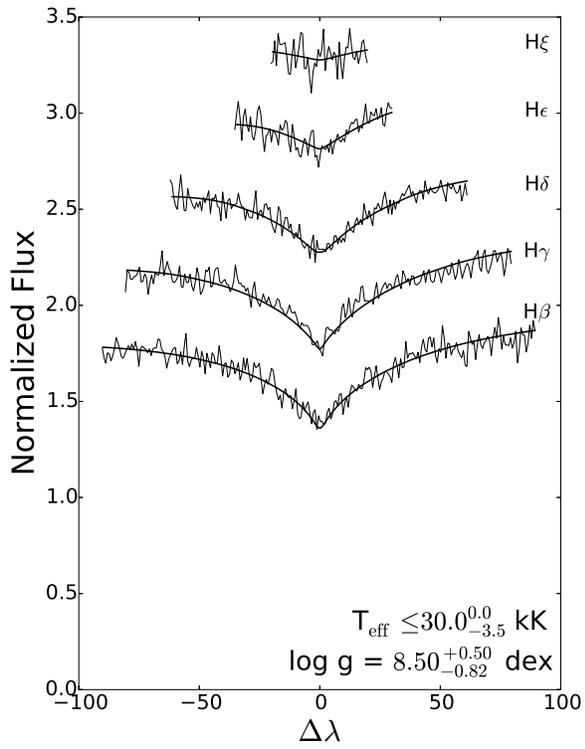}
    \end{subfigure}
    ~  
    \begin{subfigure}[]{}
        \includegraphics[width=0.47\textwidth]{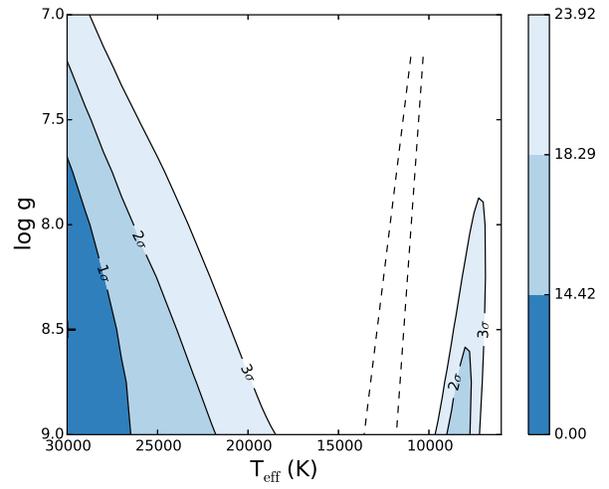}
    \end{subfigure}
    \caption{Same as Fig.~\ref{fig:19010plots}, but for KIC 7346018.}\label{fig:blue1903plots}
\end{figure}

\begin{figure}
    \begin{subfigure}[]{}
        \includegraphics[width=0.46\textwidth]{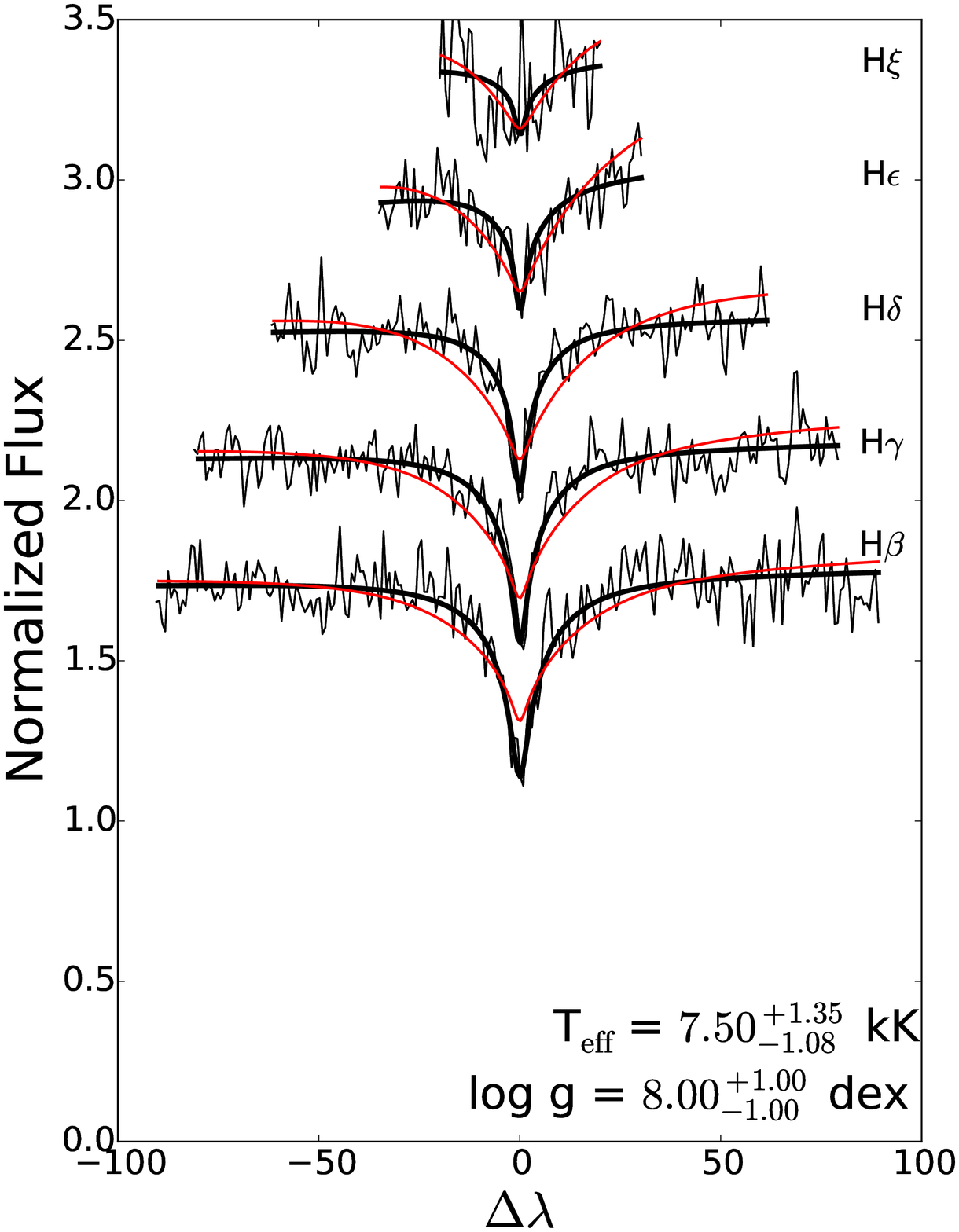}
    \end{subfigure}
    ~  
    \begin{subfigure}[]{}
        \includegraphics[width=0.47\textwidth]{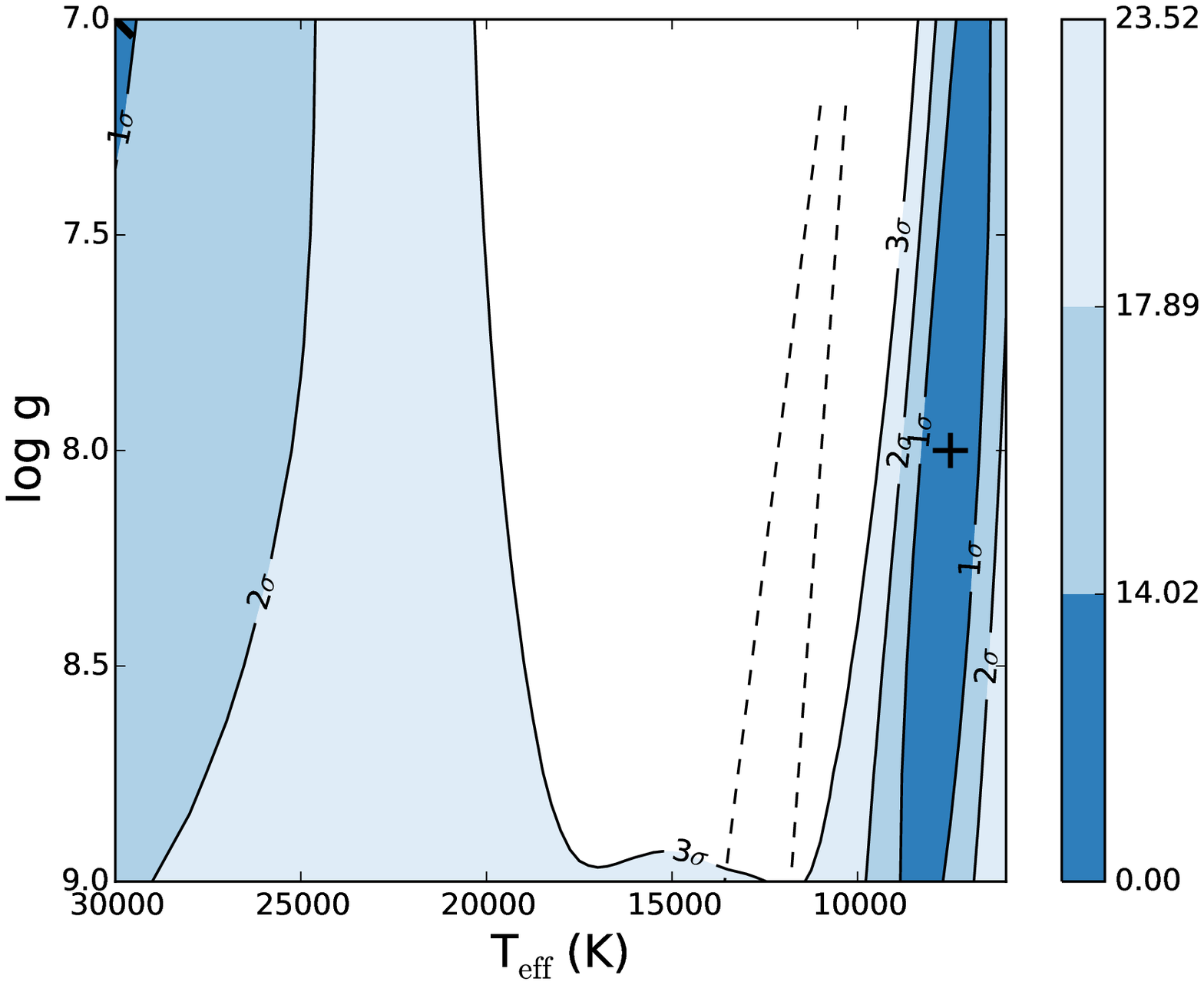}
    \end{subfigure}
    \caption{Same as Fig.~\ref{fig:19010plots}, but for KIC 8612751.  The red line fit in (a) and the $\times$ in (b) correspond to the secondary $\chi^2$ minimum of \teff\ = 30.0 kK and log $g$ = 7.00 dex.  }\label{fig:19060plots}
\end{figure}

\begin{figure}
    \begin{subfigure}[]{}
        \includegraphics[width=0.46\textwidth]{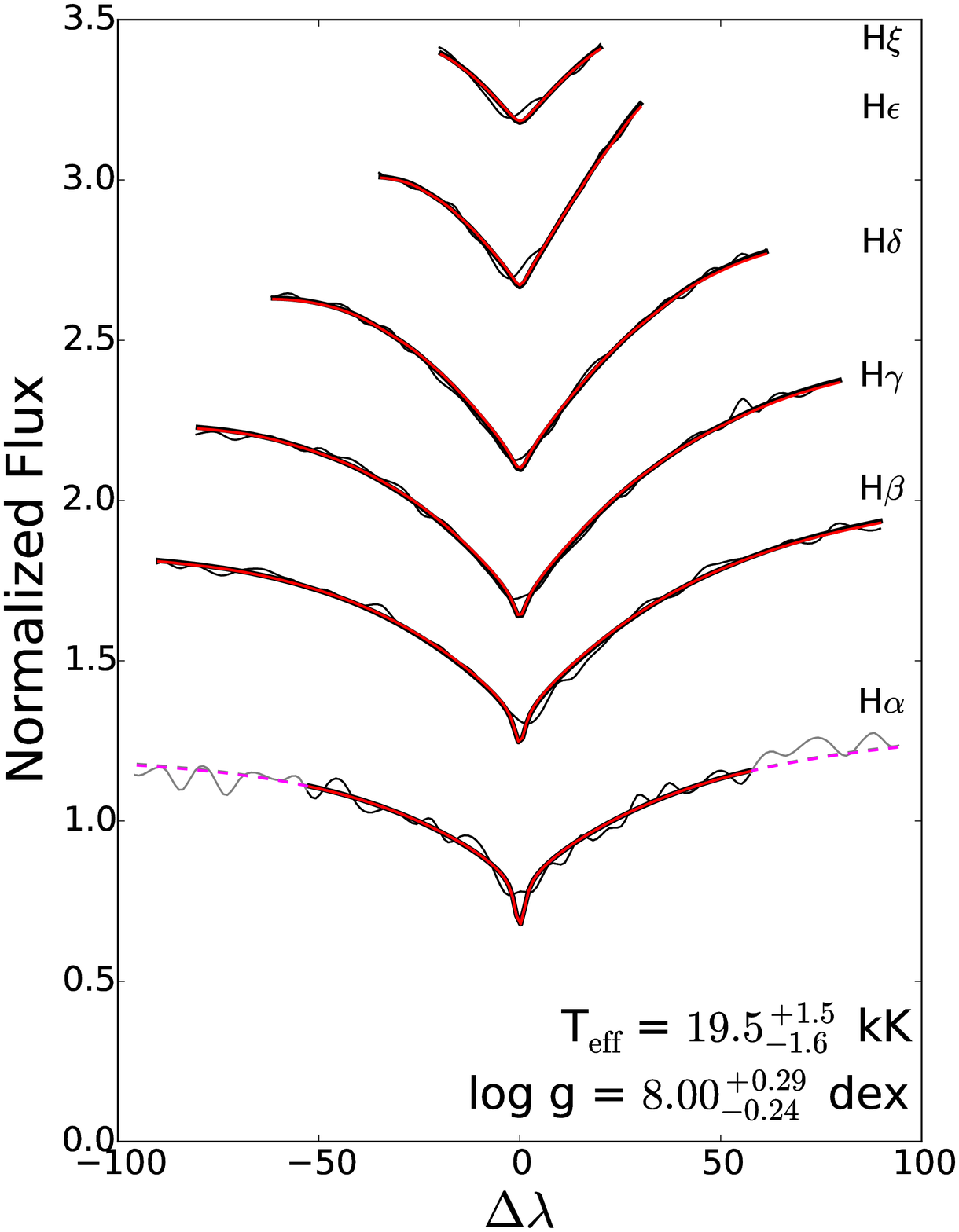}
    \end{subfigure}
    ~  
    \begin{subfigure}[]{}
        \includegraphics[width=0.47\textwidth]{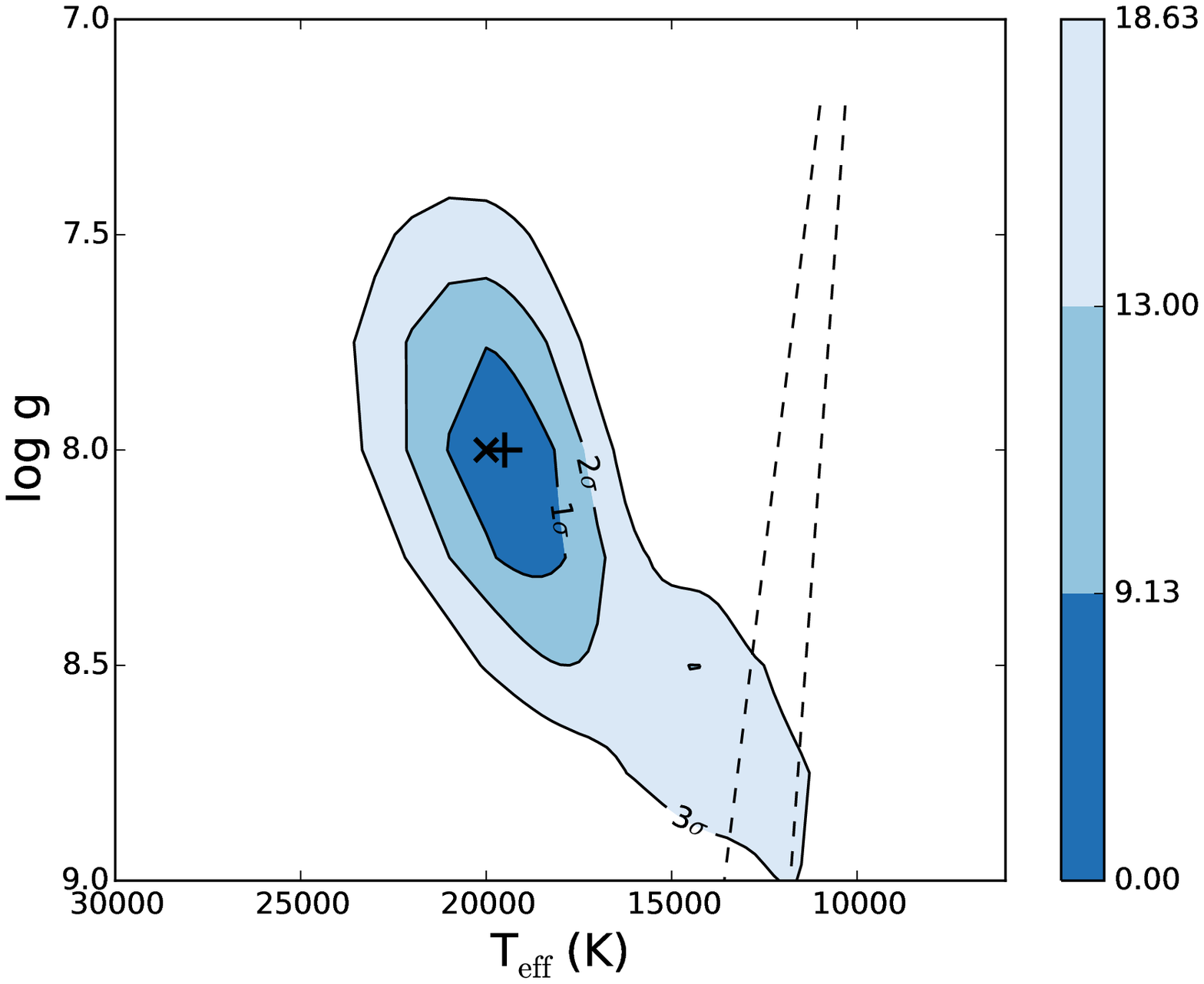}
    \end{subfigure}
    \caption{Same as Fig.~\ref{fig:18486plots}, but for KIC 4829241.  The red line fit in (a) and the $\times$ in (b) correspond to the modeled parameters from \citet{zhao13}.  }\label{fig:hsh08aplots}
\end{figure}

\begin{figure}
    \begin{subfigure}[]{}
        \includegraphics[width=0.46\textwidth]{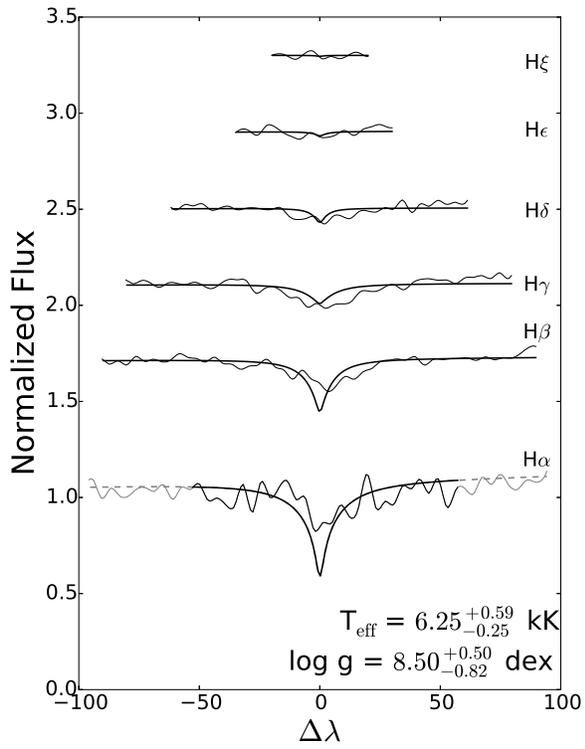}
    \end{subfigure}
    ~ 
    \begin{subfigure}[]{}
        \includegraphics[width=0.47\textwidth]{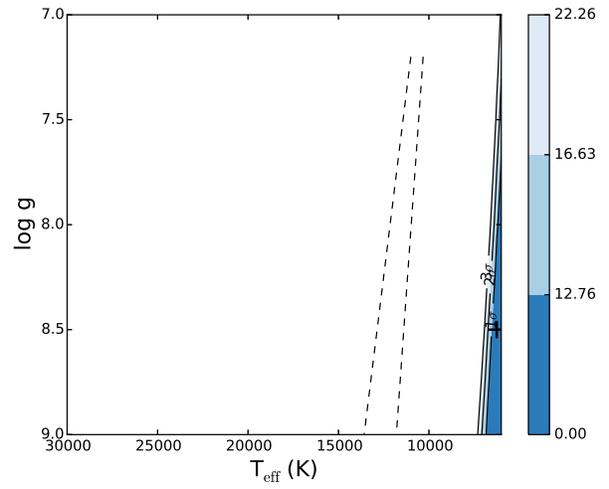}
    \end{subfigure}
    \caption{Same as Fig.~\ref{fig:18486plots}, but for KIC 6212123.}\label{fig:hsh24plots}
\end{figure}

\begin{figure}
    \begin{subfigure}[]{}
        \includegraphics[width=0.46\textwidth]{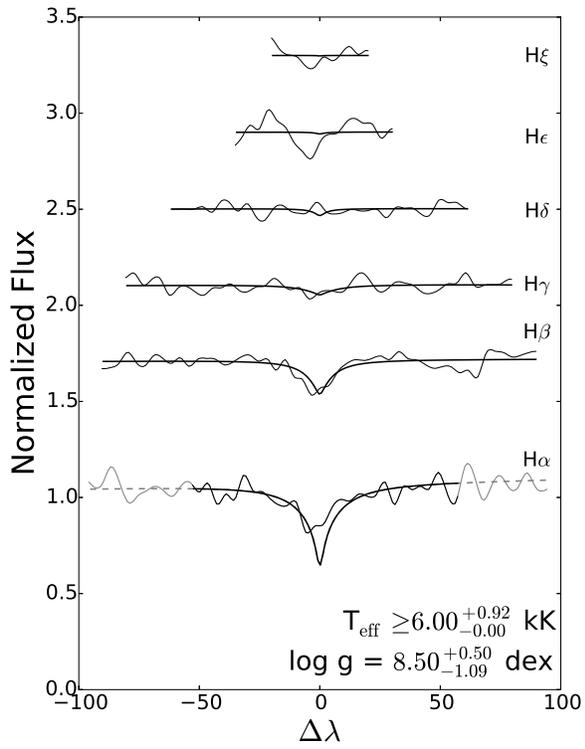}
    \end{subfigure}
    ~ 
    \begin{subfigure}[]{}
        \includegraphics[width=0.47\textwidth]{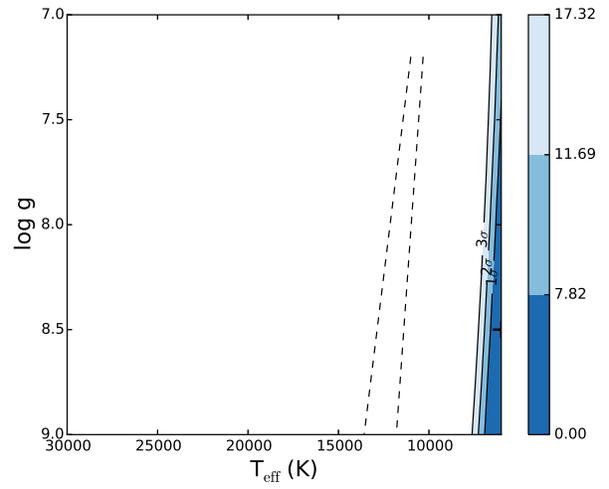}
    \end{subfigure}
    \caption{Same as Fig.~\ref{fig:18486plots}, but for KIC 3354599.}\label{fig:hsh32plots}
\end{figure}

\begin{figure}
    \begin{subfigure}[]{}
        \includegraphics[width=0.46\textwidth]{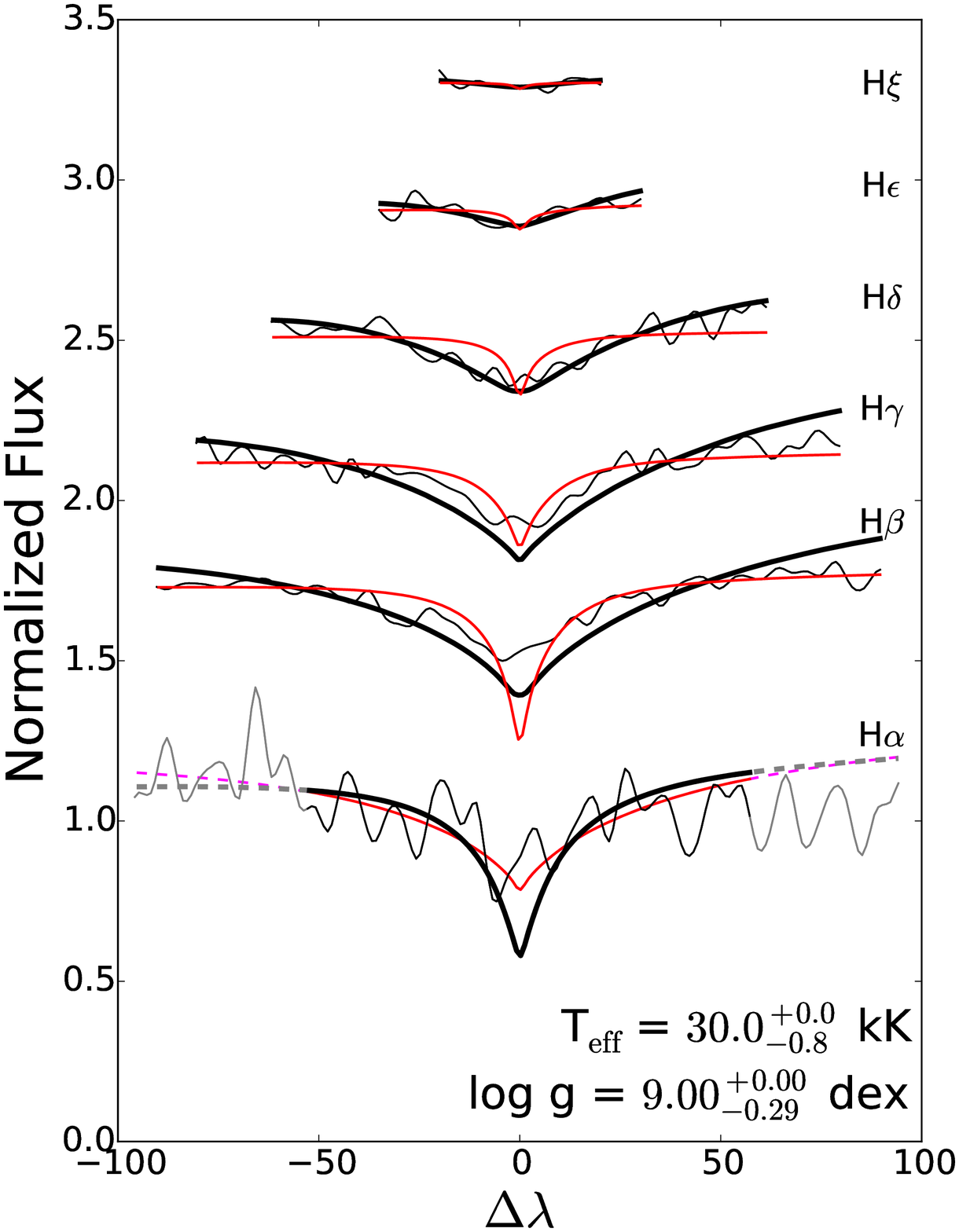}
    \end{subfigure}
    ~ 
        \begin{subfigure}[]{}
        \includegraphics[width=0.47\textwidth]{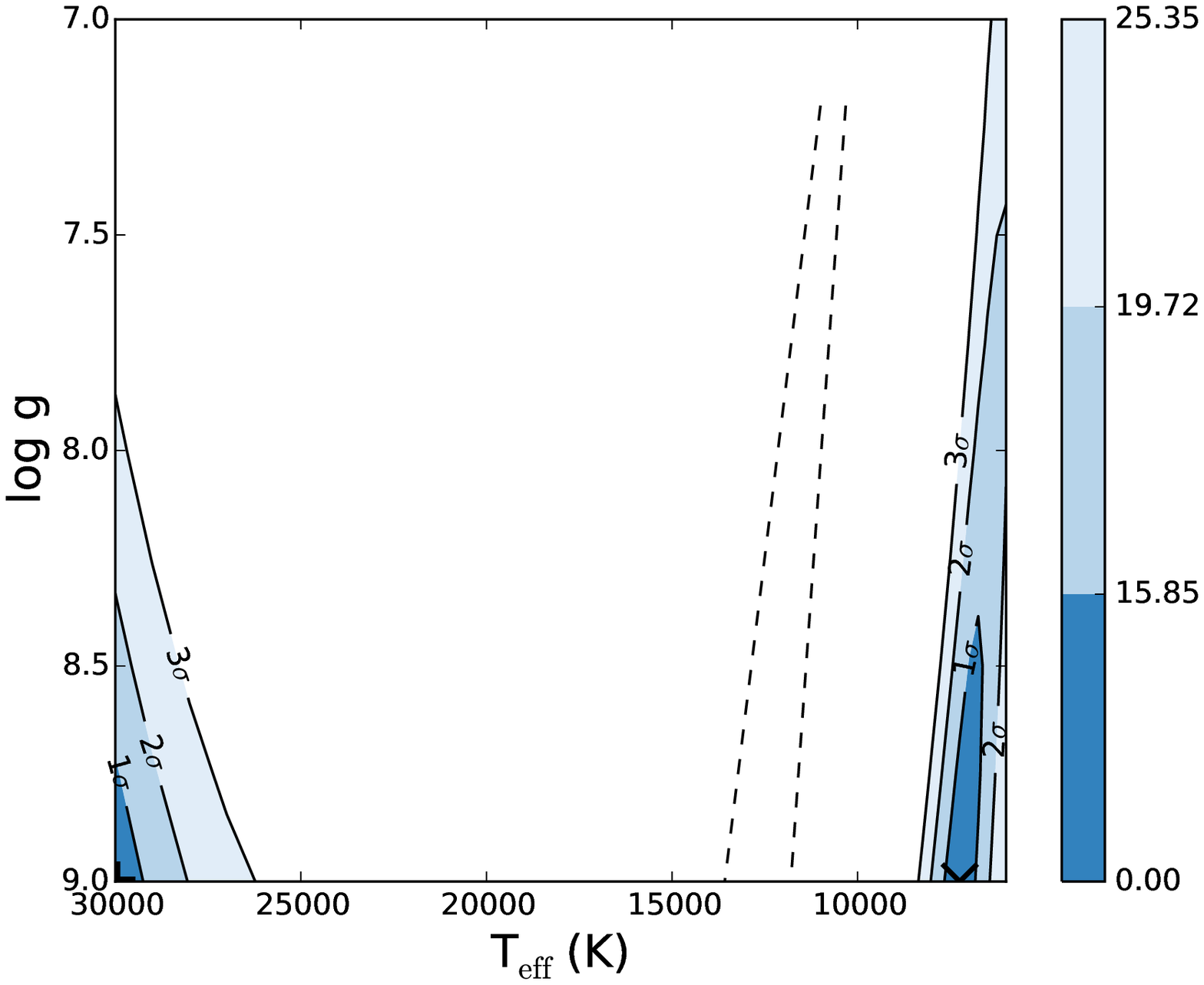}
    \end{subfigure}
    \caption{Same as Fig.~\ref{fig:18486plots}, but for KIC 10149875.  The red line fit in (a) and the $\times$ in (b) correspond to the other $\chi^2$ minimum of \teff\ = 7.25 kK and log $g$ = 9.00 dex.  }\label{fig:hsh36plots}
\end{figure}

\end{document}